\documentclass[journal]{IEEEtran}
\usepackage{subfig}
\usepackage{amsmath,amssymb,amsfonts,amsthm}
\usepackage{xcolor}
\usepackage{cite}
\usepackage{graphicx} 
\usepackage{algorithmic,algorithm}
\usepackage{extarrows}
\usepackage{comment}
\usepackage{subfig}
\usepackage[acronym]{glossaries}
\usepackage{float}

\ifCLASSINFOpdf

\else

\fi

\begin{document}

\title{Statistical Quality Comparison of the Bitstrings Generated by a Physical Unclonable Function across Xilinx, Altera and Microsemi Devices}

\author{Jenilee~Jao,~\IEEEmembership{Member,~IEEE,}
        Kristi~Hoffman,
        Brent~Emery,
        Cheryl~Reid,
        Ryan~Thomson,
        Michael~Thompson,
        Jim~Plusquellic,~\IEEEmembership{Member,~IEEE}

\thanks{J. Jao and J. Plusquellic are with the Department
of Electrical and Computer Engineering, University of New Mexico, Albuquerque,
NM, 87131 USA e-mail: jimp@ece.unm.edu.}
\thanks{K. Hoffman, B. Emery, R. Thomson, C. Reid and M. Thompson are with Collins Aerospace, RTX}
\thanks{Manuscript received June 10, 2024}}

\markboth{Transactions on Computer-Aided Design,~Vol.~x, No.~y, Month~Year}%
{Shell \MakeLowercase{\textit{et al.}}: Bare Demo of IEEEtran.cls for Journals}

\maketitle

\begin{abstract}%
Entropy or randomness represents a foundational security property in security-related operations, such as key generation. Key generation in turn is central to security protocols such as authentication and encryption. Physical unclonable functions (PUF) are hardware-based primitives that can serve as key generation engines in modern microelectronic devices and applications. PUFs derive entropy from manufacturing variations that exist naturally within and across otherwise identical copies of a device. However, the levels of random variations that represent entropy, which are strongly correlated to the quality of the PUF-generated bitstrings, vary from one manufacturer to another. In this paper, we evaluate entropy across a set of devices manufactured by three mainstream FPGA vendors,  Xilinx, Altera and Microsemi. The devices selected for evaluation are considered low-end commercial devices to make the analysis relevant to IoT applications. The SiRF PUF is used in the evaluation, and is constructed nearly identically across the three vendor devices, setting aside minor differences that exist in certain logic element primitives used within the PUF architecture, and which have only a minor impact on our comparative analysis. The SiRF PUF uses a high-resolution time-to-digital converter (TDC) crafted from high-speed carry-chain logic embedded within each device to measure path delays in an engineered netlist of logic gates as a source of entropy. Therefore, our analysis includes an evaluation of actual path delay variation as it exists across the three device classes, as well as a statistical evaluation of the PUF-generated bitstrings. A reliablity analysis is also provided using data collected in industrial-standard temperature experiments to round out the evaluation of important statistical properties of the PUF.

\end{abstract}

\begin{IEEEkeywords}
FPGAs, Physical Unclonable Functions, Entropy Analysis.
\end{IEEEkeywords}

\IEEEpeerreviewmaketitle

\section{Introduction}\label{Section:Introduction}

\IEEEPARstart{P}{hysical} Unclonable Functions (PUFs) are secure alternatives to storing secret keys in expensive, secure non-volatile memories (NVMs). PUFs leverage entropy or random variations that occur unavoidably in the fabrication processes associated with modern microelectronic device manufacturing. Physical layer variations which occur in transistor gate, source and drain geometries, in contact and via resistances, in the widths of wires, and in transistor threshold voltages, manifest as variations in the electrical parameters of the transistors and gates which implement a digital circuit. The most important, and most significantly affected, are parameters that impact the delay of signals propagating through circuit netlists that implement digital functions. Given this rich source of entropy, many types of PUF architectures have been proposed that leverage delay variations as the primary source of entropy available for key and authentication bitstring generation. 

In this paper, we investigate the level of entropy available to the ShIft-register Reconvergent-Fanout (SiRF) PUF when implemented on three different low-cost FPGA-SoC device classes, namely, the Zynq 7010 SoC device manufactured by Xilinx, the CycloneV SoC device manufactured by Altera and the PolarFire SoC device manufactured by Microsemi. Propagation delays through logic gates within SiRF's engineered netlist are measured using a high resolution time-to-digital converter (TDC) instantiated in the programmable logic (PL) of each SoC device. Our analysis isolates delay components introduced by within-die variations by applying data post-processing methods designed to remove global chip-to-chip and environmentally-induced variations from the measured path delays. We present results that illustrate the level of within-die variations using TDC-measured values of the actual delays, as well as the stability of these delay variations across twenty-five instances of the devices, and across a range of temperatures from $-40^oC$ to $85^0C$. We refer to the delay variations introduced by changes in environmental conditions as temperature-voltage noise (TV-noise), despite the fact that we did not vary supply voltage in our experiments.

The SiRF PUF algorithm is used to post-process the TDC-measured delay values into reproducible bitstrings. Statistical tests are applied to measure the statistical quality of the bitstrings, with assessments performed to determine the level of uniqueness and reliability, as well as a suite of tests for measuring randomness. The statistical tests utilize Hamming distance to measure uniqueness and reliability, and the NIST statistical test suite for evaluating randomness \cite{NIST2010}. Entropy and min-entropy are also reported for completeness. The statistical quality of the generated bitstrings for each of the device classes are compared to evaluate the impact of the FPGA fabric primitives, interconnect components and manufacturing technology on the level of entropy and noise. An entropy(signal)-to-(TV-)noise (SNR) ratio is derived which reflects a critically important overall statistical quality metric for each of the device classes.

The specific contributions of this work include:

\begin{itemize}
\item An analysis of entropy and TV-noise across multiple copies of SoC FPGAs manufactured by three mainstream manufacturers using the SiRF PUF architecture, with the entropy source designed nearly identically within the programmable logic associated with each device class.
\item An instantiation of a time-to-digital-converter (TDC) on each of the device classes for obtaining high-resolution measurements of path delays, and a description of the implementation challenges and differences.
\item A statistical quality assessment of the bitstrings produced by a set of devices from each device class, a comparison of important statistical quality metrics, namely uniqueness, reliability and randomness, and the formulation of a SNR metric that reflects that overall statistical quality of the PUF-generated bitstrings.
\end{itemize}

The remainder of this paper is organized as follows. Section \ref{Section:RelatedWork} discusses related work. Section \ref{Section:SystemOverview} describes the experimental designs, including differences in the implementations within each device class. Section \ref{Section:ExperimentalResults} presents experimental results, while Section \ref{Section:Conclusions} presents our conclusions.

\section{Related Work} \label{Section:RelatedWork}

The work presented in \cite{Gaj2012} report RO PUF bitstring statistics for Xilinx, Altera and Microsemi devices as we do in this paper. However, the work was done on small numbers of devices fabricated in older technology nodes, in particular, 13 Altera Cyclone II, 5 Xilinx Spartan 3 and 5 Actel Fusion FPGAs, and across a limited temperature range of $30~^\circ\text{C}$ to $80~^\circ\text{C}$. Moreover, the paper does not carry out an analysis of PUF soft data, e.g., actual RO counts, to determine the ratio of entropy-to-TV-noise, nor does it provide a full statistical assessment of the bitstrings across commercial-grade environmental conditions.

A more recent study uses the TERO-PUF on a Xilinx Spartan 6 in 45 nm technology and an Altera CycloneV in 28 nm \cite{Marchand2018}. Although larger sets of devices are used (30 Spartan 6 and 18 CycloneV devices), the size of the bitstrings analyzed is very small at 128-bits, and the reliability assessment is carried out over a limited range between $-15~^\circ\text{C}$ to $65~^\circ\text{C}$ and for the Xilinx devices only. To their credit, the authors did investigate supply voltage variations, which was not possible in our study because of the large number of board modifications required, but did so only at room temperature. Last, a tolerance of 10\% is used for reliability, which restricts the results of the analysis to fuzzy-match-based authentication, and not encryption keys, unless error correction is used. 

An analysis of chip-to-chip, within-die and TV-noise variations in a set of 512 ring-oscillators (ROs) instantiated on 125 Xilinx Spartan 3E FPGAs is presented in \cite{Maiti2010}. Although the study is focused on one device class, it presents an analysis of RO frequency variation, a.k.a. an analysis of RO soft data. The authors of \cite{Wilde2014} expand on the analysis performed in \cite{Maiti2010} by applying normality and similarity tests, principle component analysis and entropy estimation to the RO data sets. In \cite{Maes2013}, the author investigates an accurate reliability model for PUFs, which assumes error probabilities are not uniform across all PUF cells, and derives a heterogeneous model as an alternative to commonly used fixed error rate models. 

A soft data-based thresholding scheme is proposed in \cite{Mai2014} that utilizes an error avoidance methodology, similar to the methodology proposed in \cite{Ju2012}. The authors of \cite{Guilley2018} describe a signal(entropy)-to-(TV-)noise ratio (SNR) similar to the one applied empirically in our work, but the analysis is applied to RO and Loop PUFs. More recently in \cite{Mexis2023}, a simulation-based framework is proposed that estimates the reliability of response bits, and which can be used to filter unreliable bits.


Unlike previous work, the FPGA-SoCs used in this work possess the same feature size, which enables a better apples-to-apples comparison. In particular, the Zynq 7010 is manufactured using TSMC's 28HPL process \cite{Zynq7000TechNode2015} \cite{Tambara2015}, the CycloneV is manufactured on TSMC's 28LP process \cite{CycloneVTechNode2015} and the PolarFire is manufactured on UMC's 28 nm SONOS process \cite{PolarFireTechNode2017}. The core power supply voltages are 1.0 V, 1.1 V and 1.0 V, respectively. A second important contribution of this work is the derivation of a entropy(signal)-to-(TV)noise (SNR) ratio for each device class. The SNR ratio is fundamental to predicting the overall quality of the PUF architecture and its generated bitstrings, as we will show.

\section{System Overview} \label{Section:SystemOverview}

\begin{figure*}[ht]
    \centering
    \includegraphics[width=7.0in,keepaspectratio=true]{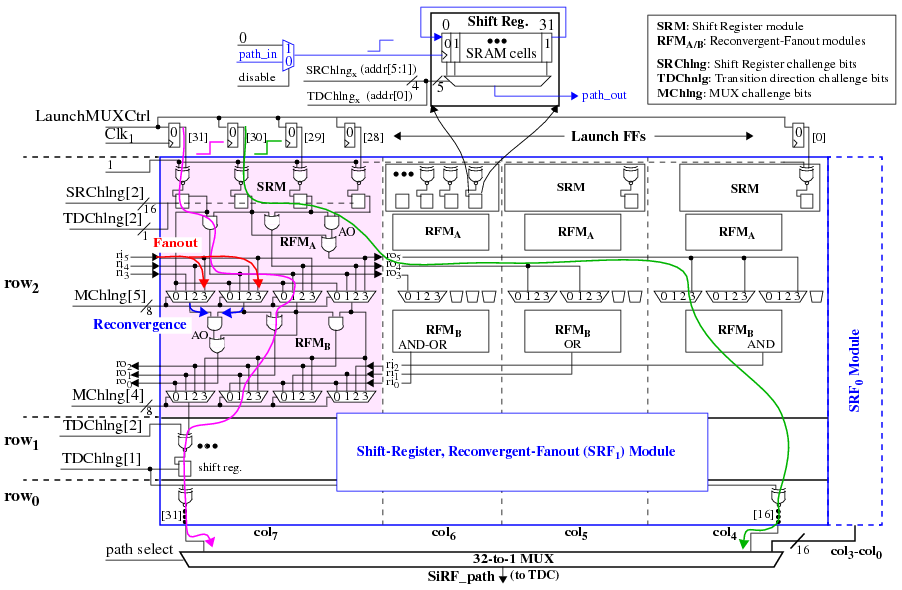}
    \caption{SiRF block diagram highlighting multiple, simultaneous signal path propagations and an instance of reconvergent-fanout.}
    \label{fig:SiRF_block_diagram}
    \vspace{-10pt}
\end{figure*}

In this section, we describe the implementation details of the SiRF PUF for each of the three device classes, as well as the differences that exist in the specific logic gate primitives available in the device technology libraries. 

The SiRF PUF architecture is shown as a block level diagram in Fig. \ref{fig:SiRF_block_diagram}. The architecture is modular, constructed as a set of interconnected blocks arranged in rows and columns. The example architecture shown in the figure, and used in the experiments on the devices in each of the device classes, is composed of three rows, $row_0$ through $row_2$, and eight columns, $col_0$ through $col_7$. The Launch FFs shown along the top of the figure launch signal transitions into the netlist components which traverse successive rows of shift-registers, logic gates and MUXes. Two signal transition paths are illustrated in the figure, which show signals moving from top to bottom and left to right. Signal paths can also wrap around either edge of the module using the rotate inputs $ri_x$ and outputs $ro_y$, creating a complex, diverse network of paths through the module. \textcolor{black}{No place and route constraints are needed or used during the implementation of the SiRF PUF, except as noted below for the implementation of the TDCs on the three device architectures.}


The netlist is engineered to remain glitch-free, ensuring that exactly one transition propagates along any given signal path. The glitch-free characteristic of the netlist is critical to obtaining reliable measurements of path delays, especially when operating the PUF under extreme environmental conditions. Each row can be configured with challenge bits to propagate either rising or falling edges, but not both. Therefore, the entropy associated with both transitions can be combined using a challenge which controls the transition direction bits ($TDClng[x]$) shown on the left side of the figure with arbitrary assignments of '0' for falling and '1' for rising transitions. Glitch-free operation is guaranteed by forcing all transitions to be either rising or falling within any given row and by using only non-inverting logic gates within the network.

Other components of the challenge control which path is selected through the shift-registers, labeled $SRChlng[x]$, and which paths through the 4-to-1 MUXes are selected to drive the next row, labeled $MChlng[y]$. Each module includes a set of XNOR gates that invert falling transitions that may be generated by the previous row to ensure that the shift-registers are capable of continuing signal propagation. From the callout shown along the top of Fig. \ref{fig:SiRF_block_diagram}, the incoming signals to a module drive the clock signal of the shift registers, where only rising transitions will cause the shift registers to shift the bit sequence by one bit position to the right. The shift-registers are initialized with an alternating sequence of '0's and '1's, which ensures any 1-bit shift will create either a rising or falling transition on the output of the shift-register. 

The netlist is also engineered to create a large number of instances of reconvergent-fanout. The left side of Fig. \ref{fig:SiRF_block_diagram} illustrates the concept of reconvergent-fanout using the rotate-in signal $ri_5$. A rising transition propagating from upstream nodes drive $ri_5$, which fans out to the inputs labeled 3 on two of the 4-to-1 MUXs shown by the red arrows. Assuming the challenge $MChlng_a$ is set such that both of these inputs are selected, the MUX outputs reconverge on the inputs of the AND gate. If a rising transition is propagating, then the signal that arrives last along one of the two branches will dominate the timing on the AND gate output, i.e., the AND gate output will not switch from low to high until both rising edges have arrived. Given that proprietary vendor place \& route tools create the implementation of the SiRF netlist without constraints, it is unknown which branch has a physically longer path, e.g., longer wire lengths, without inspecting the layout. It is also possible that both branches of the reconvergent-fanout are nearly equal in delay. In either case, there is uncertainty regarding which path dominates the timing, which complicates model-building techniques that require physical layer models. Moreover, for the equal delay case, it may happen that the branch which dominates the timing varies from one device to another, further increasing the level of uncertainty.

All paths through the nelist eventually emerge and connect to a 32-to-1 MUX shown along the bottom of Fig. \ref{fig:SiRF_block_diagram}. The timing engine state machine logic controls the path select bits of the 32-to-1 MUX, enabling each of the signal paths to be directed to a time-to-digital converter (TDC) (discussed below). 

\subsection{Xilinx, Altera and Microsemi Implementation Details} \label{Section:DeviceImplementationDetails}

We describe differences in the logic element primitives amongst the three FPGA-SoC device classes in this section. The TDC utilizes hardwired carry chain primitives, which have different underlying structures in the programmable fabrics of each device class. The shift register primitives are also implemented differently. Zynq devices support a 32-bit shift register primitive while Cyclone and PolarFire, to the best of our knowledge, infer shift registers from RTL behavioral descriptions rather than providing device primitives or hard macros.

\subsubsection{TDC}

\begin{figure*}[ht]
    \centering
    \includegraphics[width=5.5in,keepaspectratio=true]{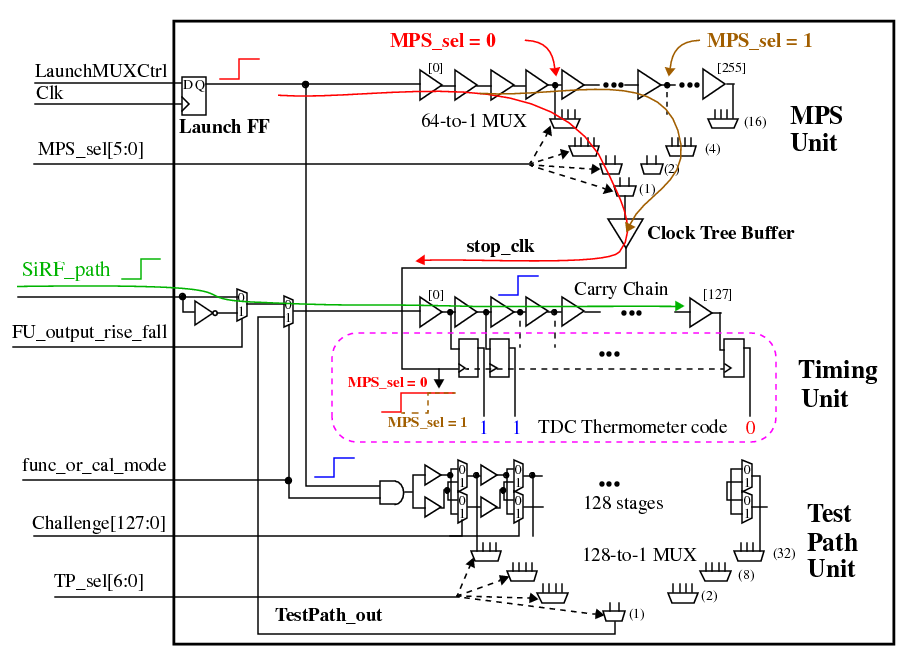}
    \caption{Schematic diagrams showing the Major Phase Shift (MPS), Timing, and Test Path elements of the TDC. }
    \label{fig:SiRF_PUF_TDC}
    \vspace{-10pt}
\end{figure*}

A block diagram of the TDC is shown in Fig. \ref{fig:SiRF_PUF_TDC}. The TDC is composed of three submodules, called the \textit{Major Phase Shift Unit} (MPS), the \textit{Timing Unit} and the \textit{Test Path Unit}. The Timing Unit is constructed using hard-wired carry chain components which makes it possible to measure path delays with a resolution in the 10's of picoseconds range. Carry chains are commonly embedded as primitives in FPGA PL-side architectures to enable CAD tools to optimize timing during the synthesis of RTL code. Addition and subtraction are very common functional unit operations in the control and/or data paths of RTL code and the embedded carry chains are leveraged to improve their performance. For the TDC, the high speed propagation capability along the carry chain, and the ability to connect the outputs of the carry chain buffers to FFs, provides a mechanism to obtain timing resolution of path delays that are on order of 10X better than what is possible using equivalent LUT-based resources.

A timing measurement is performed using a launch-capture strategy, where the system clock (Clk) driving the Launch FFs in Fig. \ref{fig:SiRF_block_diagram}, and the Launch FF in Fig. \ref{fig:SiRF_PUF_TDC}, is used to launch a rising transition into the SiRF PUF netlist and MPS Unit simultaneously. The rising edge propagates through the SiRF PUF netlist to the 32-to-1 MUX and drives the \textit{SiRF\_path} input of the TDC, while the MPS Unit edge propagates along the delay chain to a selected tap point. The simultaneous launching of both signals creates a race condition, with the signal propagating through the MPS Unit serving as a \textit{stop\_clk} signal that halts the race. The relative delays of both signals determines how far the rising edge \textit{SiRF\_path} signal propagates along the carry chain before the MPS Unit signal asserts the clock inputs to the Timing Unit FFs. After the stop clock event, the Timing Unit FFs store a \textit{thermometer code}, i.e., a sequence of 1's followed by a sequence of 0's. The number of 0's in the thermometer code (TC) represents a digital delay value (\textbf{DV}) for the tested path, which is then stored by the SiRF PUF algorithm in a block RAM (BRAM).

The MPS Unit incorporates a MUXing structure to enable the selection of a tap point. During testing of a path, a state machine repeats the launch-capture test with incrementally larger tap point selections, where each increment increases the delay of the \textit{stop\_clk} signal, until a valid TC, i.e., one with a non-zero number of 1's, is produced. Therefore, the actual delay of the tested path is the sum of the TC and the selected tap point delay. To determine the delays corresponding to the set of tap points (64 tap points are shown in Fig. \ref{fig:SiRF_PUF_TDC}), a calibration operation is carried out prior to any SiRF netlist testing operations. Calibration utilizes the Test Path Unit to configure test paths of various lengths, which are used as the test path signal to the Timing Unit, instead of the \textit{SiRF\_path} signal input. A sequence of calibration tests are performed using test paths of different lengths to enable an accurate average delay value to be computed for each tap point. The final DV stored in the BRAM is the sum of a valid TC and the calibration-derived delay of the selected tap point. Details of the calibration process are omitted here but can be found in \cite{bulletproof2018}.


\begin{figure}[ht]
    \centering
    \includegraphics[width=3.0in,keepaspectratio=true]{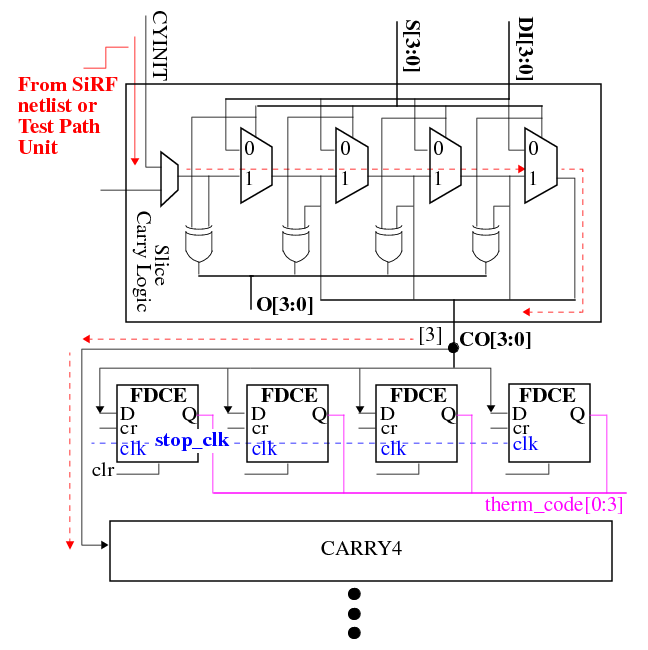}
    \caption{Zynq 7010 LUT configuration that implements the initial portion of TDC \cite{XilinxSeries7ug953}.}
    \label{fig:Zynq_TDC}
    \vspace{-10pt}
\end{figure}

As mentioned, the carry chain component is very common in FPGA PL-side architectures, and is used in the configuration of the TDCs in all three of the FPGA device classes. The layout details differ from one device class to another, but nearly the same level of resolution is achievable. 

The Zynq 7 series device class provides a CARRY4 primitive (upgraded to a CARRY8 for UltraScale+ architectures) for implementing fast carry chains. The TDC in the Zynq 7010 device is configured to use 32 copies of the CARRY4 block connected in series, to define a carry chain of length 128. The first copy of the CARRY4 chain is shown in Fig. \ref{fig:Zynq_TDC}, where the path-under-test (PUT) drives the \textit{CYINIT} input of the topmost CARRY4. A set of thermometer code FFs within the SLICE are connected to the CO (carry-out) outputs of the CARRY4, and the carry-out[3] signal is routed to the carry-in of the next CARRY4 block. The \textit{stop\_clk} signal is derived from a global clock buffer, which drives the clock inputs of the thermometer code (TC) FFs.

\begin{figure}[ht]
    \centering
    \includegraphics[width=3.4in,keepaspectratio=true]{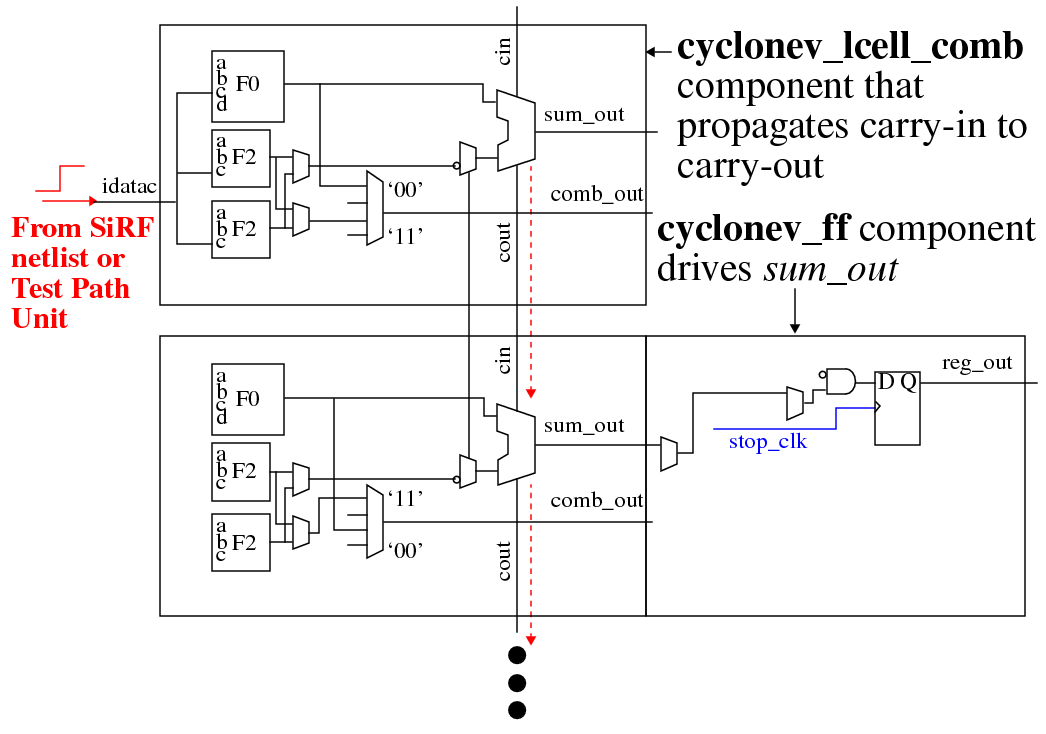}
    \caption{CycloneV ALM configuration that implements the initial portion of TDC \cite{AlteraALM}.}
    \label{fig:CycloneV_TDC}
    \vspace{-10pt}
\end{figure}

A carry chain is instantiated in the CycloneV devices using the \textit{cyclonev\_lcell\_comb} library component, as shown in Fig. \ref{fig:CycloneV_TDC} (thanks goes to \cite{Charlot2021} for the solution). A sequence of Altera FPGA adaptive logic modules (ALMs) are shown, which define the first two elements of the TDC. The top-most ALM is used to introduce the PUT edge into the carry chain. Unlike the Zynq device, TC FFs are connected to the \textit{SUM\_OUT} outputs of the LCELL primitive within the same ALM. The carry chain is constructed with 256 elements, in contrast to the Zynq implementation, which contains 128 elements. The carry chain length only impacts the speed of TDC calibration process, and does not effect the timing resolution of the TDC. Therefore, differences in the length of the TDC are inconsequential to the analysis presented herein.

\begin{figure}[ht]
    \centering
    \includegraphics[width=3.4in,keepaspectratio=true]{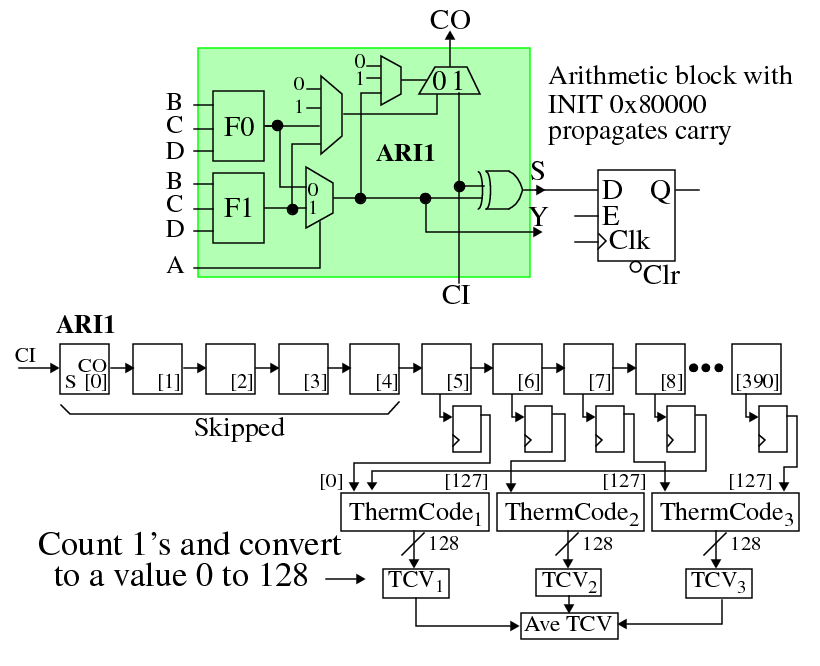}
    \caption{PolarFire LUT configuration that implements the initial portion of TDC \cite{PolarFireARI1}.}
    \label{fig:PolarFire_TDC}
    \vspace{-10pt}
\end{figure}

PolarFire defines an ARI1 primitive that can be used to implement a fast carry chain, as shown in Fig. \ref{fig:PolarFire_TDC}. Unlike the Zynq and Cyclone devices, the delay through each of the carry chain elements, defined as a sequence of ARI1 primitives, is not monotonically increasing, which creates 'holes' in the TCs, i.e. '0's in the sequence of '1's. However, from timing simulation, we found that sequences defined using every third ARI1 element are monotonic. Therefore, a set of three 128-bit TC chains are created by connecting every third element in a sequence as shown in the figure. Moreover, we also determined that the first 5 elements of the TDC carry chain were not well correlated with the remaining values, and are therefore skipped as shown in the figure. The length of the carry chain is expanded to 391 elements to accommodate these constraints.

The three TCVs obtained for a PUT in the PolarFire devices are averaged using the expression in Eq. \ref{Eq:AveTCV}, which expands the range of the TDC from 128 to 192. As we will see, this pseudo-averaging of three TCs per test reduces measurement noise levels over the single-valued TDCs implemented on the Zynq and Cyclone devices. However, the SiRF PUF algorithm enables multiple TC samples of each PUT to be collected and averaged, which is used in the Zynq and Cyclone analyses to make the comparison of noise levels nearly equivalent.

\begin{equation}
    \text{TCV}_{\text{Ave}} = (TCV_1 + TCV_2 + TCV_3)/2
    \label{Eq:AveTCV}
\end{equation}

\subsubsection{Shift Registers}

\begin{figure}[ht]
    \centering
    \includegraphics[width=3.4in,keepaspectratio=true]{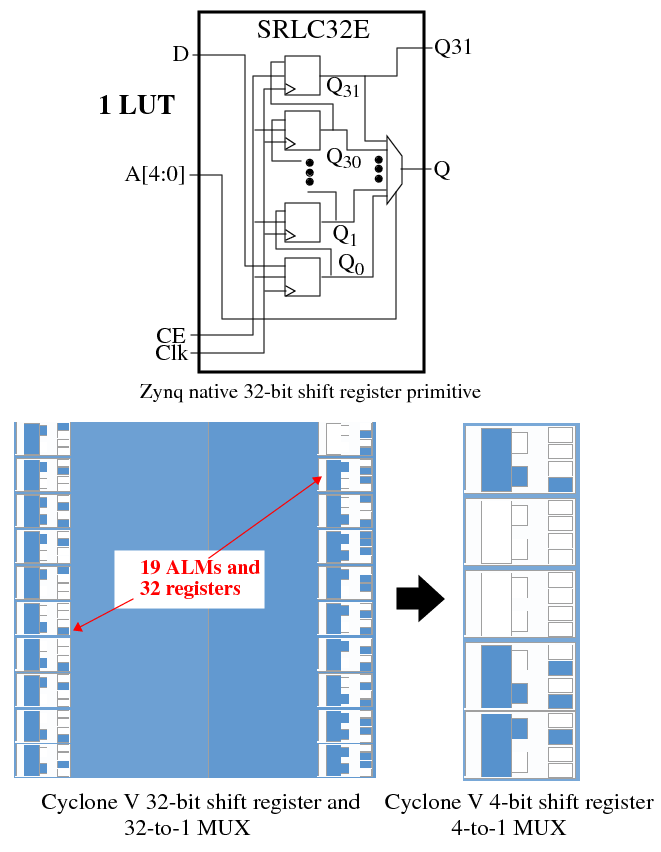}
    \caption{Implementation of a 32-bit shift registers on the Zynq 7010 (top) \cite{XilinxSeries7ug953} CycloneV (bottom-left), and a 4-bit shift register for the CycloneV (bottom-right).}
    \label{fig:Zynq_CycloneV_ShiftReg_MUX}
    \vspace{-10pt}
\end{figure}

Native device primitive support for shift registers exists only on the Zynq device, as a unisim library component called the SRLC32E. The schematic for the SRLC32E is shown along the top of Fig. \ref{fig:Zynq_CycloneV_ShiftReg_MUX}, and its implementation uses one LUT. The Zynq primitive uses the LUT resources to implement both the shift register and selection MUX because the circuit structures required to implement the LUT are nearly equivalent to the circuit structures required for the shift-register-MUX combination. 

In contrast, we were not able to find a Cyclone and PolarFire dedicated shift register-MUX primitive, and instead, construct the functionality using multiple look-up table primitives. The lower left portion of Fig. \ref{fig:Zynq_CycloneV_ShiftReg_MUX} shows the layout of an equivalent 32-bit shift-register-MUX combination on the Cyclone. The fabric resources needed include 19 ALMs and 32 FFs. Although not shown, PolarFire requirements are similar. In an attempt to match the number of resources used for the SiRF PUF netlist across all device classes, the Cyclone and PolarFire shift-register-MUX implementations are reduced from 32 bits to 4 bits, as shown on the right-bottom side of Fig. \ref{fig:Zynq_CycloneV_ShiftReg_MUX}. This ensures the path lengths are similar in all three implementations, which in turn, improves the fairness of the comparisons of entropy, TV-noise and the bitstring metrics. 

\begin{figure*}[ht]
    \centering
    \includegraphics[width=7.0in,keepaspectratio=true]{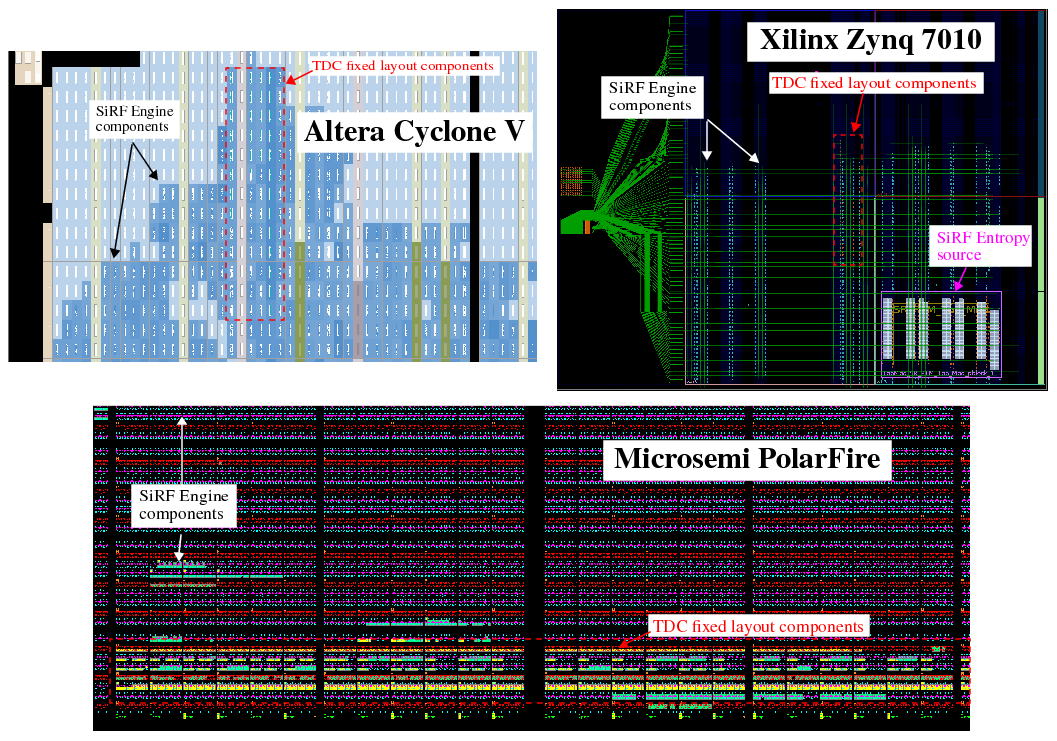}
    \caption{Implementation views of SiRF PUF on the three device classes, with highlighted TDC components.}
    \label{fig:Zynq_Cyclone_PolarFire_layouts}
    \vspace{-10pt}
\end{figure*}

\subsection{Architecture} \label{Section:Arch}

Portions of the implementation views of the SiRF PUF on the Zynq 7010, CycloneV and PolarFire devices are shown in Fig. \ref{fig:Zynq_Cyclone_PolarFire_layouts}. The red-dotted rectangles highlight the regions corresponding to the fixed components of the TDC implementations. As indicated earlier, the carry chains of the TDCs in the Zynq, Cyclone and PolarFire are 128, 256 and 390 elements in length, respectively. The PL fabric resources in all three devices easily accommodate the integration of the TDCs. 

All three designs were synthesized with a timing constraint of 50 MHz, and all three produced SiRF netlist path delay values in the range of 5 ns to 20 ns. The carry chains in the TDC implementations support path delay measurements in the range of 2 to 4 ns. Therefore, the delay range expansion provided by the MPS Unit and the calibration process described earlier are essential to enabling SiRF netlist path delays to be measured. 




\section{Experimental Results} \label{Section:ExperimentalResults}

\begin{figure*}[ht]
    \centering
    \includegraphics[width=6.0in,keepaspectratio=true]{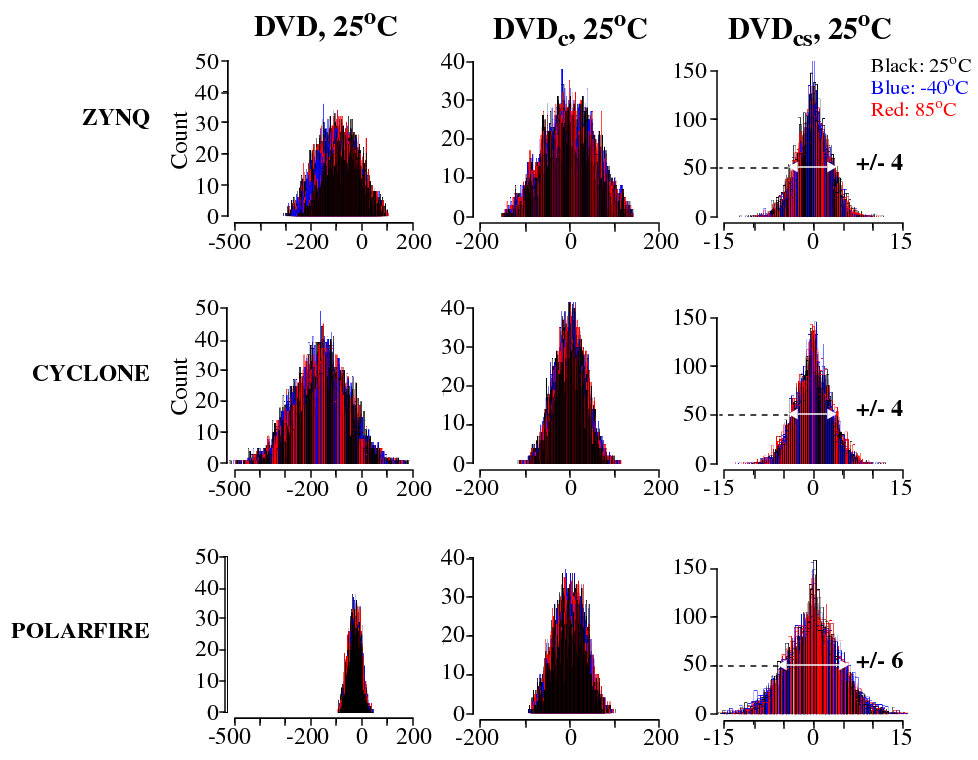}
    \caption{Superimposed distributions of 2048 $DVD$, $DVD_c$ and $DVD_{co}$ from 25 Zynq, Cyclone and PolarFire devices illustrating the SiRF PUF group processing operations.}
    \label{fig:ZCP_DVD_DVDc_DVDco}
    \vspace{-10pt}
\end{figure*}

The major objective of our analysis is to measure and compare the average level of entropy and TV-noise present in the SiRF netlist path delays across the three device classes. The evaluation is carried out using a set of 25 devices from each device class. The same set of characterization vectors are applied to all 75 devices, and a set of 64,000 high-resolution delay values (DVs) are collected from each device under nominal conditions. We refer to this data as the \textbf{enrollment data}. Five additional \textbf{regeneration} experiments are carried out which repeat these experiments at temperatures given by \{$-40~^\circ\text{C},  0~^\circ\text{C}, 25~^\circ\text{C}, 50~^\circ\text{C}, 85~^\circ\text{C}$\}. The combined enrollment and regeneration data sets are used for the entropy and TV-noise analyses, while the bitstring analyses uses the enrollment data and regeneration data in the traditional way for evaluation of reliability and other statistical metrics. 

As we have done in previous works \cite{Plusquellic:2022}, we post-process the DVs to compensate for global process variations and changes in environmental conditions as a means of extracting delay variations introduced by within-die process variations. The first section of the results shows the effect of applying our proposed mathematical transformations to the raw DVs to accomplish this goal, which also reveals the levels of TV-noise that remain. The SiRF PUF's entropy-TV characterization process selects a subset of the 64,000 DVs that are best described as \textit{compatible}, where compatibility is defined as path delays that scale approximately linearly with changes in temperature conditions. We provide a brief description of the entropy-TV characterization process in this paper, and refer readers to \cite{Plusquellic:2022} for a detailed description of the process. The values reported for the average level of entropy and TV-noise are derived using only the DV-compatibility sets.

The next section of our results focuses on a quantitative evaluation of overall levels of entropy and TV-noise for each device class. The applied data transformations remove global biases from the raw DV from each device class to enable a comparison of the signal(entropy)-to-TV-noise ratios. The last section presents results of a statistical analysis of the bitstrings from each device class, including analysis of randomness, uniqueness and reliability. Parameters to the SiRF PUF algorithm's reliability enhancement techniques are tuned for each device class to make the comparison as fair as possible.

\subsection{DV Post-Processing}

The SiRF PUF algorithm applies a sequence of transformations to a set of 2048 rising DVs (DVR) and 2048 falling DVs (DVF). The superimposed distributions generated by operations important to our analysis in this paper are shown in Fig. \ref{fig:ZCP_DVD_DVDc_DVDco} for sets of 25 devices from the Zynq, Cyclone and PolarFire device classes. The following summarizes the operations that produce these distributions. 
\begin{enumerate}
    \item The DVDiffs module creates a one-to-one pairing relationship between the 2048 DVR and 2048 DVF stored in BRAM, and subtracts the DVF from the DVR to produce DVD. The superimposed distributions from the 25 devices in each device class are shown in the left column of Fig. \ref{fig:ZCP_DVD_DVDc_DVDco}.
    \item The global process and environmental variation (GPEV) module applies a pair of linear transformations to the DVD to produce $DVD_c$, as shown in the center column of the figure. GPEV removes delay variations introduced by chip-to-chip (global) process variations, and significantly reduces temperature-supply voltage effects on the path delays. 
    \item The SpreadFactors module eliminates path length bias effects, which are present because the paths through the SiRF netlist vary in length. This occurs because no placement or routing constraints are used to fix the positions of the gates and wires in the SiRF netlist. The right-most column in the figure depicts distributions of $DVD_{cs}$.
\end{enumerate}

\begin{figure*}[ht]
    \centering
    \includegraphics[width=7.0in,keepaspectratio=true]{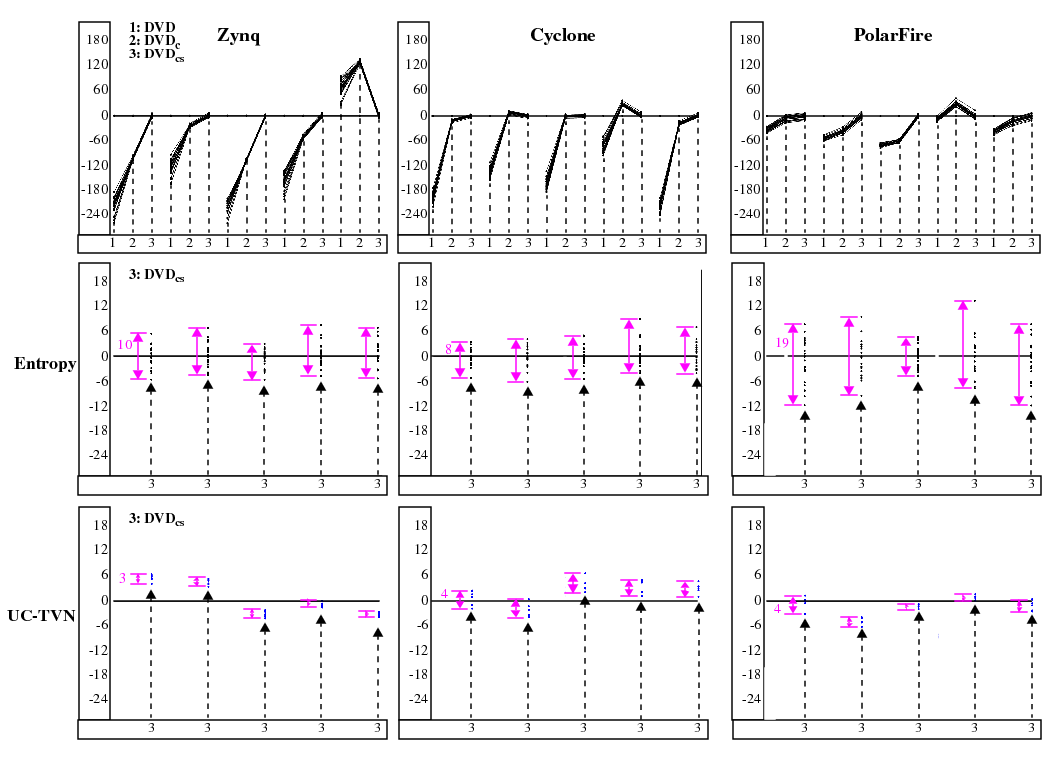}
    \caption{Example $DVD$, $DVD_c$ and $DVD_{co}$ from 25 Zynq, Cyclone and PolarFire devices illustrating entropy and TV-noise assessment.}
    \label{fig:ZCP_DVD_DVDc_DVDco_and_TVNose}
    \vspace{-10pt}
\end{figure*}

Several characteristics are revealed in the distributions. First, the DVD distributions associated with the Zynq device class exhibit shifts left-and-right that are not as dramatic in the Cyclone and PolarFire distributions. These shifts are introduced by chip-to-chip (global) process variations. The PolarFire distributions are nearly coincident, exhibiting very little global process variation effects. Unfortunately, wafer-lot information is not available for the device sets, which might explain the disparity observed across the device classes. Second, the compensation carried out by GPEV yields wider distributions for the Zynq devices, which suggests that larger differences exist in the rising and falling delays of these devices, especially when compared with the narrow distributions associated with the PolarFire devices. And third, the widths of the $DVD_{cs}$ distributions are nearly the same for the Zynq and Cyclone device classes, while the PolarFire distributions are approximately 33\% wider. The $DVD_{cs}$ distributions portray the level of entropy available to the PUF, and therefore, the PolarFire devices dominate this metric.

The level of entropy is critically important to all PUF architectures but cannot be assessed without considering the level of TV-noise present. Entropy below the noise floor cannot be accessed by the PUF unless error correction methods are utilized during bitstring generation. The SiRF PUF, however, utilizes error avoidance methods which require the level of entropy to be above the TV-noise floor. 

\subsection{Analysis of Entropy and TV-Noise} \label{Subsection:EntropyAndTVNoise}

A key component to the assessment of quality of the SiRF PUF on each of the three device classes is to evaluate the ratio of entropy-to-TV-noise (SNR). The graphs in Fig. \ref{fig:ZCP_DVD_DVDc_DVDco_and_TVNose} provide a visual aid to how the SNR is computed. The top-most row shows the first five $DVD$, $DVD_c$ and $DVD_{cs}$ (of the 2048) from the 25 superimposed distributions in Fig. \ref{fig:ZCP_DVD_DVDc_DVDco}. Therefore, each group of points in a column contains 25 points, one for each device. The sequences of line-connected points show the transformations from $DVD$ through $DVD_{cs}$, which translate the points vertically toward 0.0 and reduce their vertical spread. 

The second row of graphs labeled \textbf{Entropy} zooms in by a factor 10 and shows only the $DVD_{cs}$. Both of our metrics for entropy and TV-noise are computed using these points. Entropy is computed for each group of points as the spread or range of the points around 0.0, which is annotated by the magenta lines and arrows. As an example, the level of entropy is labeled as 10, 8 and 19, respectively, for the left-most set of points of each device class. The third row of graphs shows an equivalent metric for TV-noise. Here, only the first device from the set of 25 in each device class is shown, and the points correspond to the $DVD_{cs}$ computed across the enrollment and 5 temperature (regeneration) corners. The vertical spread (range) of the points in this case represents TV-noise that was not eliminated by GPEV. We refer to this residual noise as uncompensated TV-noise or \textbf{UC-TVN}. As indicated, UC-TVN defines the noise floor, and it must be smaller than the level of entropy in order for the SiRF PUF's error avoidance scheme to be effective. As an example, the range of UC-TVN is annotated as 3, 4 and 4 respectively, for the left-most groups of points in each graph, which shows the entropy-to-UC-TVN requirement is met, i.e., UC-TVN is at least a factor of 2 smaller than entropy.

\begin{figure}[ht]
    \centering
    \includegraphics[width=3.4in,keepaspectratio=true]{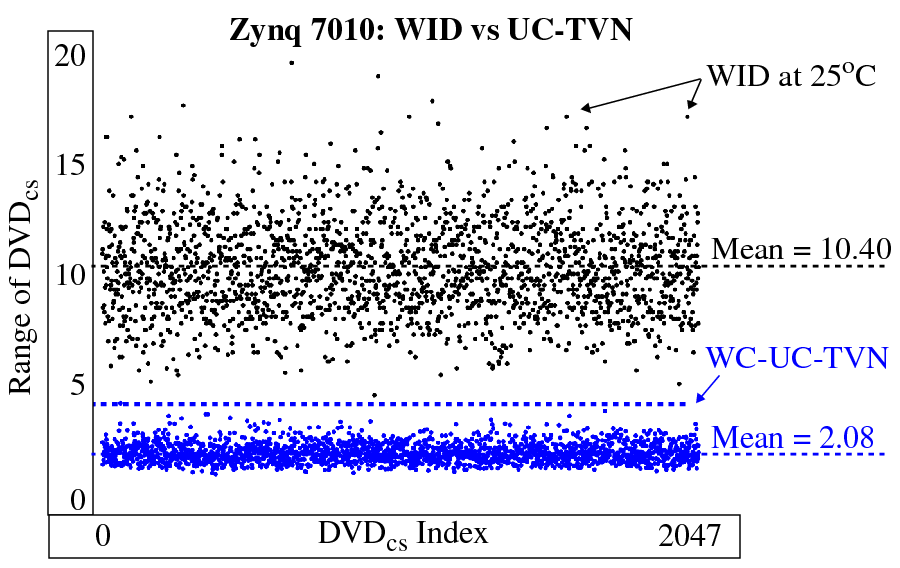}
    \caption{Zynq 7010: WID vs. UC-TVN using 2048 $DVD_{cs}$ from one challenge set.}
    \label{fig:Zynq_WID_Vs_TVN}
    \vspace{-10pt}
\end{figure}

\begin{figure}[ht]
    \centering
    \includegraphics[width=3.4in,keepaspectratio=true]{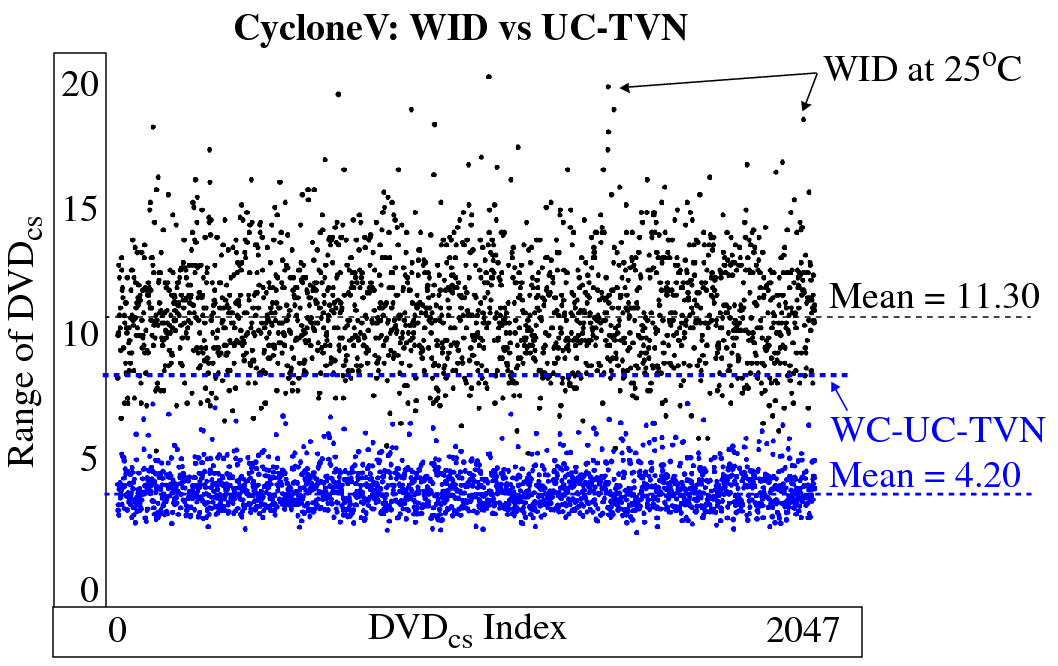}
    \caption{CycloneV: WID vs. UC-TVN using 2048 $DVD_{cs}$ from one challenge set.}
    \label{fig:Cyclone_WID_Vs_TVN}
    \vspace{-10pt}
\end{figure}

\begin{figure}[ht]
    \centering
    \includegraphics[width=3.4in,keepaspectratio=true]{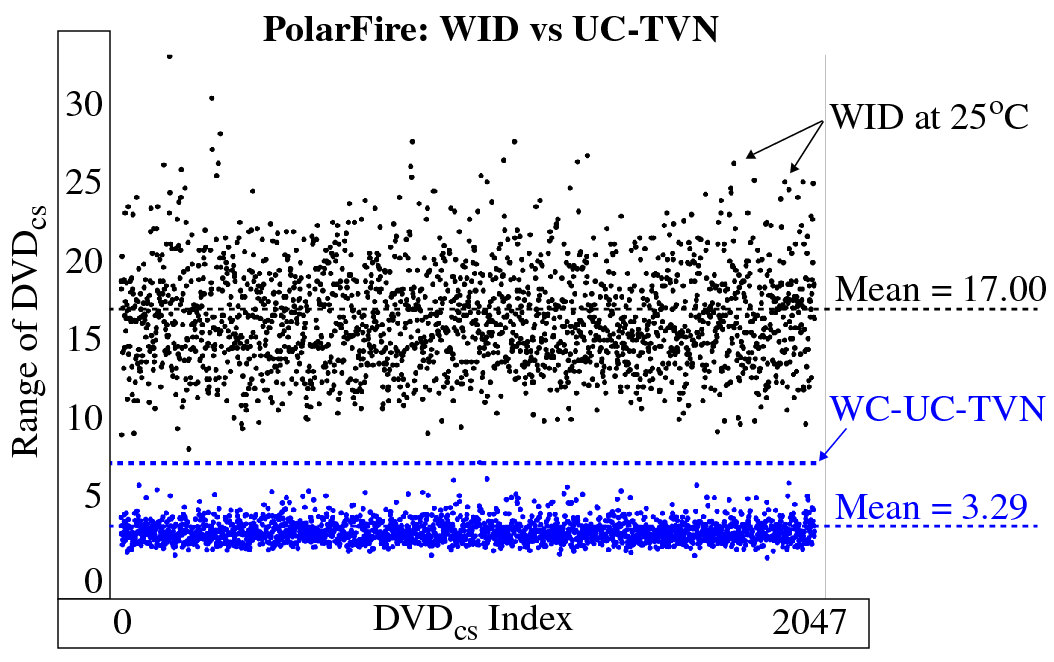}
    \caption{PolarFire: WID vs. UC-TVN using 2048 $DVD_{cs}$ from one challenge set.}
    \label{fig:PolarFire_WID_Vs_TVN}
    \vspace{-10pt}
\end{figure}

An overall assessment of entropy and UC-TVN for each of the three device classes is shown graphically in Fig. \ref{fig:Zynq_WID_Vs_TVN} through Fig. \ref{fig:PolarFire_WID_Vs_TVN}. Entropy is referred to as within-die variation or \textbf{WID} in these graphs because WID better describes what it represents. Here, we plot the WID as a set of black points and UC-TVN as a set of blue points. Each of the 2048 points in either case represent the range of the $DVD_{cs}$ across the 25 devices as described in reference to Fig. \ref{fig:ZCP_DVD_DVDc_DVDco_and_TVNose}. As indicated in the figures, WID is computed using data collected at $25~^\circ\text{C}$.

The mean values of WID and UC-TVN for all 2048 points are also shown in the three figures, and are given for Zynq as 10.40 and 2.08, for Cyclone as 11.30 and 4.20 and for PolarFire as 17.00 and 3.29. The corresponding ratios of WID-to-UC-TVN, i.e., SNR, are given as 5.0, 2.69 and 5.17, respectively, for Zynq, Cyclone and PolarFire. As is true for SNR metrics in general, the larger the ratio the better, so PolarFire is best, with Zynq as a close second, while Cyclone performs significantly worse than PolarFire and Zynq. Another metric that reveals this fact is depicted in the figures as WC-UC-TVN, which identifies the worst-case UC-TVN across all 2048 points. While the WC-UC-TVN is smaller than than the (smallest) worst-case WID for Zynq and PolarFire (a desirable characteristic), this is not the case for Cyclone. In fact there is a fair amount of overlap in the black and blue points. As we show later, the higher noise levels associated with the Cyclone device makes it more difficult to obtain our target reliability metric of 1-bit-flip-per-million.

It is difficult to speculate on why the Cyclone device class possesses higher noise levels than Zynq and PolarFire. One possibility is rooted in the layout characteristics of the programmable fabric, while another stems from the CAD tools responsible for the generation of the netlist and for placement and routing. A third possibility is related to the manufacturing facility. An analysis of the test data collected during calibration of the TDC (not included here) indicates the TDC itself is stable and is not the source of the noise. Future experiments are planned in which the SiRF netlist will be placed inside of a \textit{logic lock} region in order to fix the logic placement within the device, to determine if this improves the noise levels.

A second interesting artifact of this analysis is the different shapes of the $DVD_{cs}$ distributions shown in the right-most column of Fig. \ref{fig:ZCP_DVD_DVDc_DVDco}. As indicated earlier, Zynq and Cyclone are manufactured in a TSMC foundry, while PolarFire is manufactured by UMC. The PolarFire distribution has a wider band in the heart of the distribution (at Count = 50), while Zynq and Cyclone are narrower and very similar in shape. These characteristics might be leveraged, for example, to identify the foundry-of-origin of the device. Future work is planned to investigate this further.

\subsection{SiRF PUF Reliability Enhancement Techniques}

\begin{figure*}[ht]
    \centering
    \includegraphics[width=7.0in,keepaspectratio=true]{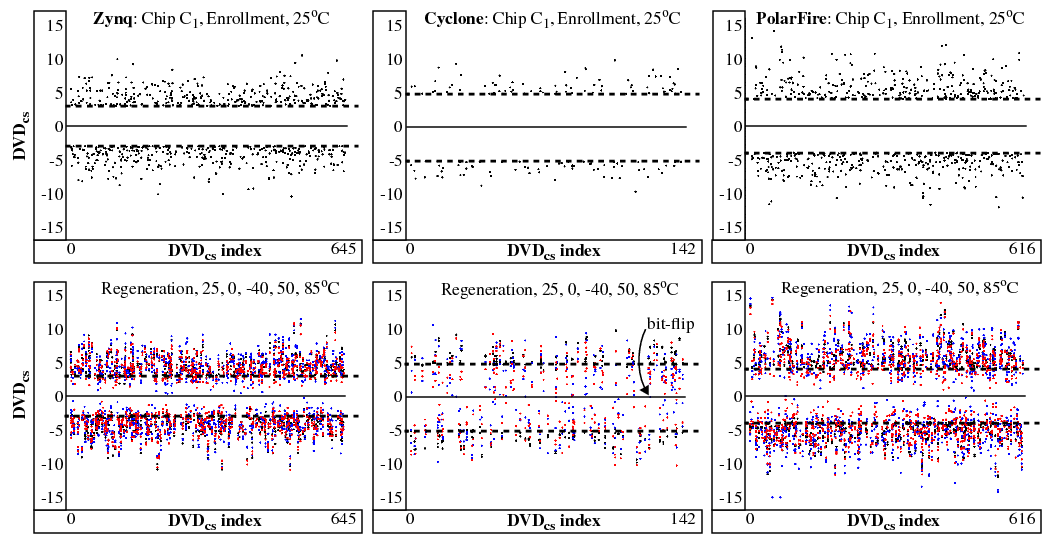}
    \caption{Illustration of bit-flip avoidance via Thresholding. Enrollment results are shown along the top row for the Zynq, Cyclone and PolarFire devices, respectively. Only $DVD_{cs}$ data points classified as strong are shown. Regeneration is shown along the bottom row, with $DVD_{cs}$ produced under different temperature conditions superimposed on the enrollment data. Encroachment of the blue (cold temperature) and red (hot temperature) data points within the threshold region illustrates the effect of UC-TVN. Data points that cross the 0 line result in bit-flip errors. }
    \label{fig:ZCP_Thresholding}
    \vspace{-10pt}
\end{figure*}


The SiRF PUF algorithm utilizes a bit-flip avoidance technique called Thresholding, in contrast to error correction, to achieve high reliability standards. Thresholding removes bits that have a high probability of flipping value during regeneration. An illustration of Thresholding is shown in Fig. \ref{fig:ZCP_Thresholding}, as it is applied to the $DVD_{cs}$ data points obtained from one each of the Zynq, Cyclone and PolarFire devices. 

Thresholding defines two symmetric thresholds around the 0 line, that are used during device enrollment to identify and eliminate unreliable bits. The top row of graphs shows thresholds of $\pm 3$, $\pm 5$ and $\pm 4$ for the Zynq, Cyclone and PolarFire devices, respectively, where $DVD_{cs}$ that fall within the threshold region are excluded (and are not shown) during enrollment. We refer to $DVD_{cs}$ that fall above the upper threshold and below the lower threshold (the points that are shown in the graphs) as strong bits in the following.

As indicated earlier, a set of challenges are applied to generate a set of 2048 $DVD_{cs}$ for each device. The number of $DVD_{cs}$ that survive the Thresholding process are given as 645, 142 and 616, for the Zynq, Cyclone, and Polarfire devices respectively. A strong bitstring, a.k.a., an encryption key, is generated from the $DVD_{cs}$ by assigning 1's to $DVD_{cs}$ that fall above the upper threshold and 0's to those that fall below the lower threshold. During enrollment, the bitstring generation algorithm also creates a helper data bitstring to record the positions of the strong bits in the sequence of 2048 $DVD_{cs}$, assigning 1 if a strong bit is generated, and 0 if a bit is skipped. The helper data does not leak information about the values of the strong bits, and can therefore be stored in non-safeguarded, standard non-volatile memory for use during regeneration. The regeneration algorithm reads and interprets the helper data bitstring, generating strong bits when helper data bits are 1.

The threshold values are determined from characterization experiments, similar to the experiments carried out here. The threshold of $\pm 5$ for the Cyclone device class is larger than the value for the Zynq and PolarFire device classes, indicating higher levels of UC-TVN. The threshold is chosen to achieve a given reliability standard, which is discussed in the next section. For a fixed level of entropy, a larger threshold reduces the number of strong bits that can be generated from the set of 2048 $DVD_{cs}$. For the example Cyclone device shown, the reduction is significant, where only 142 bits of the 2048 possible bits are classified as strong. In contrast, the Zynq device produces a strong bitstring of length 645 while the PolarFire device produces 616 strong bits. The consequence of fewer bits is the requirement to run the SiRF PUF algorithm a second time using a different set of challenges as a means of, e.g., generating a 256-bit encryption key for the Cyclone device, while only one iteration is needed for Zynq and PolarFire devices.

The second row of graphs in Fig. \ref{fig:ZCP_Thresholding} shows $DVD_{cs}$ produced by the same three devices while subjecting them to different temperature conditions. The helper data bitstrings produced during enrollment are used to select the same $DVD_{cs}$ for regeneration of the strong bitstrings. The adverse impact of UC-TVN is depicted as an encroachment of the regenerated $DVD_{cs}$ into the threshold region. The threshold is selected to minimize the probability that a regenerated $DVD_{cs}$ appears on the opposite side of the 0 line, when compared to the position of the corresponding enrollment-generated $DVD_{cs}$, which would result in a bit-flip error in the regenerated strong bitstring. Despite the larger threshold for the Cyclone device, several regenerated $DVD_{cs}$ (colored blue and red) get very close to, and in one case cross, the bit-flip line. The Zynq device performs best with respect to minimizing UC-TVN because only a threshold of $\pm 3$ is required to achieve zero bit-flip errors. The PolarFire device ranks second with a requirement of $\pm 4$ for the threshold, while the Cyclone device ranks last.

Despite the reliability enhancements provided by DV-compatibility set selection, GPEV and Thresholding, bit-flip errors can still occur. A third reliability enhancement scheme, called XMR, can be layered on top of these methods to improve reliability even further. XMR uses redundancy to encode super-strong bits from a sequence of strong bits, and adds protection against bit-flip errors by allowing, e.g., one strong bit in a sequence of three strong bits to flip value during regeneration. A correct, error-free super-strong bit is generated in these cases because majority vote is used to determine the final value of the bit during regeneration. Similar to Thresholding, the level of protection against bit-flip errors can be tuned using a parameter to XMR, where increasing the level of redundancy, e.g., from 3 to 5, 7, etc, provides higher levels of reliability \cite{Plusquellic:2022}.

\subsection{Bitstring Statistical Analysis}

\begin{figure*}[!t]
   \centering
   \subfloat[Entropy]{\includegraphics[width=3.5in]{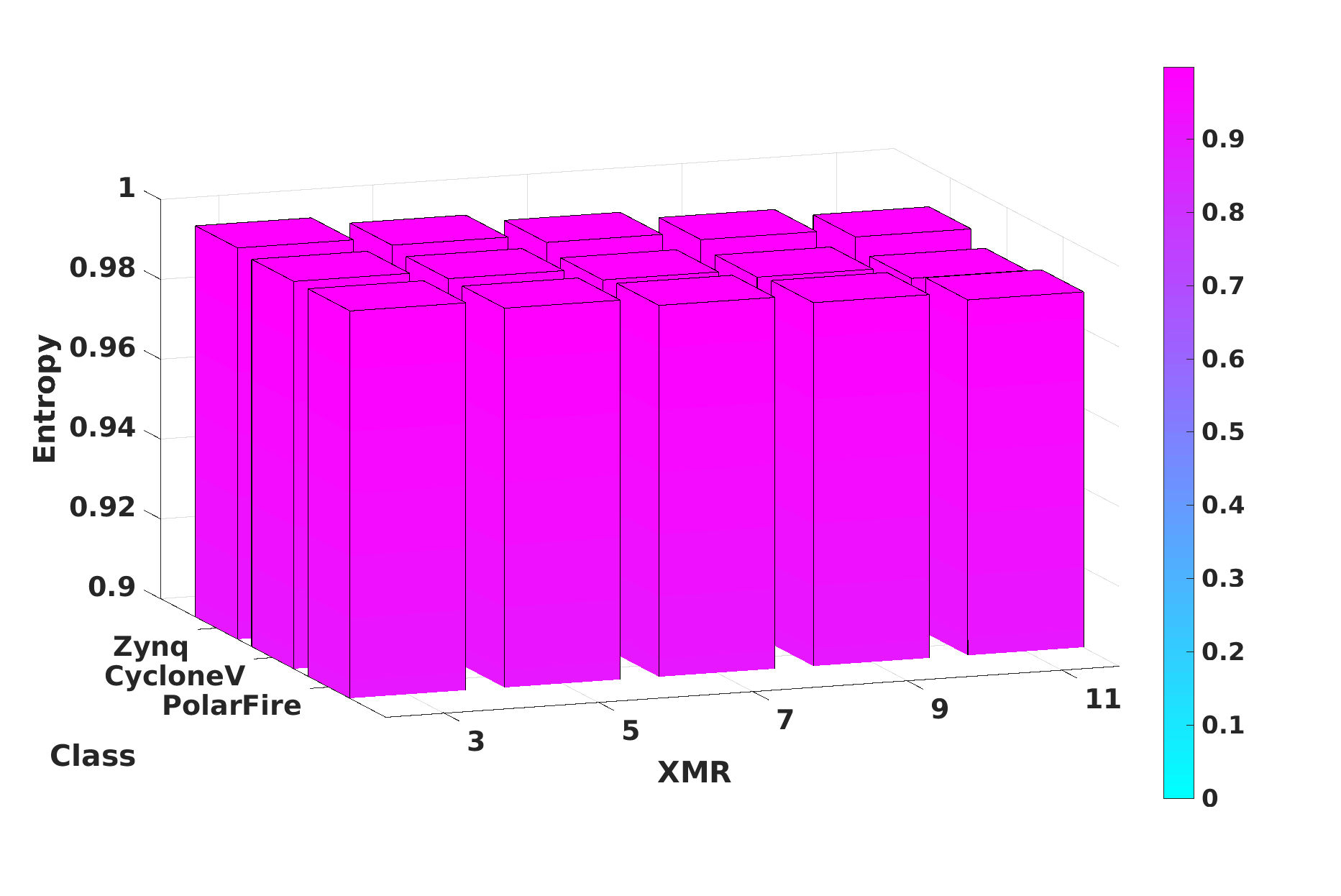}
      \label{Fig:Entropy}}
      \hfil
   \subfloat[MinEntropy]{\includegraphics[width=3.5in]{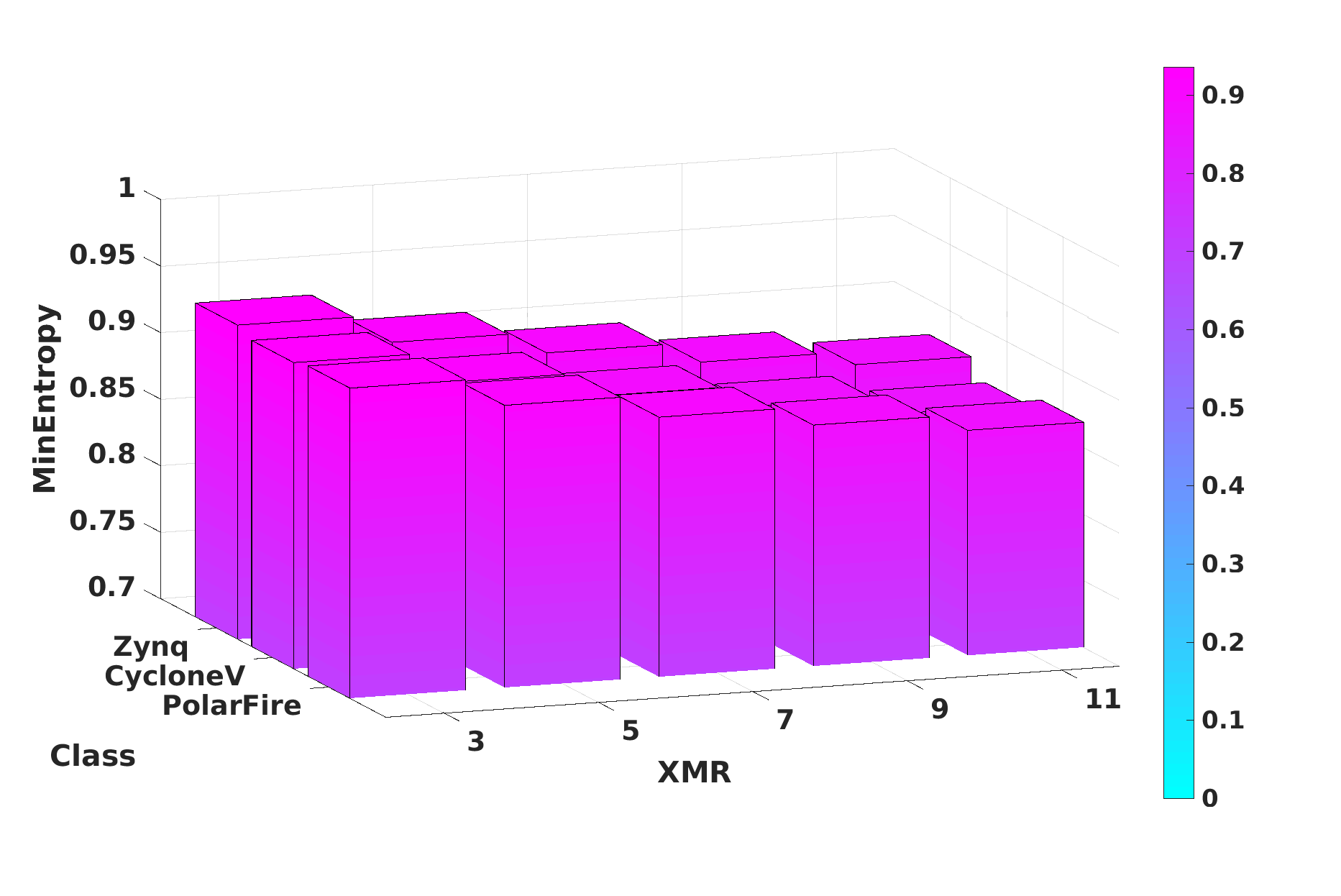}%
      \label{Fig:MinEntropy}}
   \caption{Entropy and min-entropy statistics for all device classes.}
   \label{Fig:EntropyMinEntropy}
   \vspace{-15pt}
\end{figure*}

\begin{figure*}[!t]
   \centering
   \subfloat[InterChip HD]{\includegraphics[width=3.5in]{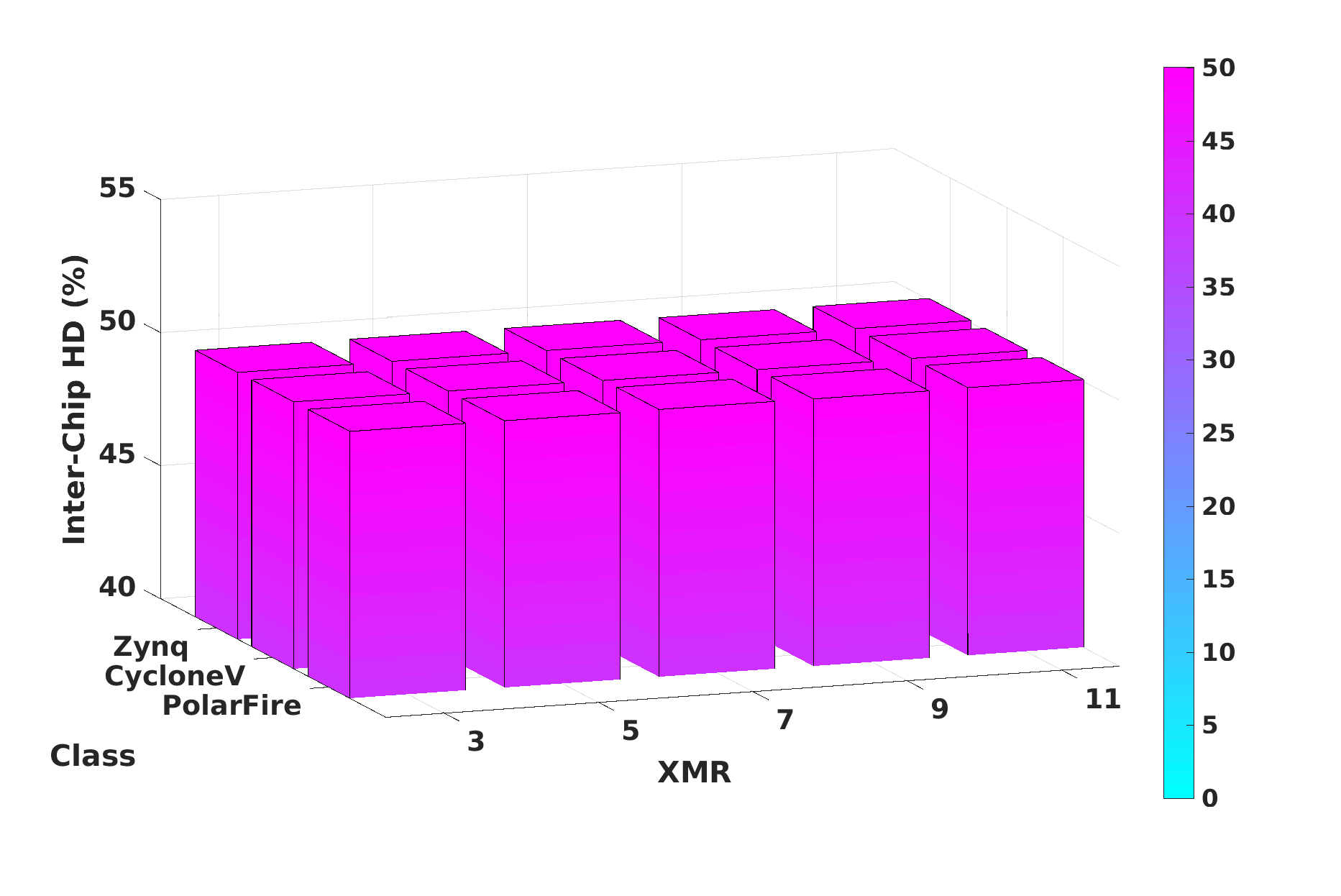}
      \label{Fig:InterChipHD}}
      \hfil
   \subfloat[Aligned InterChip HD]{\includegraphics[width=3.5in]{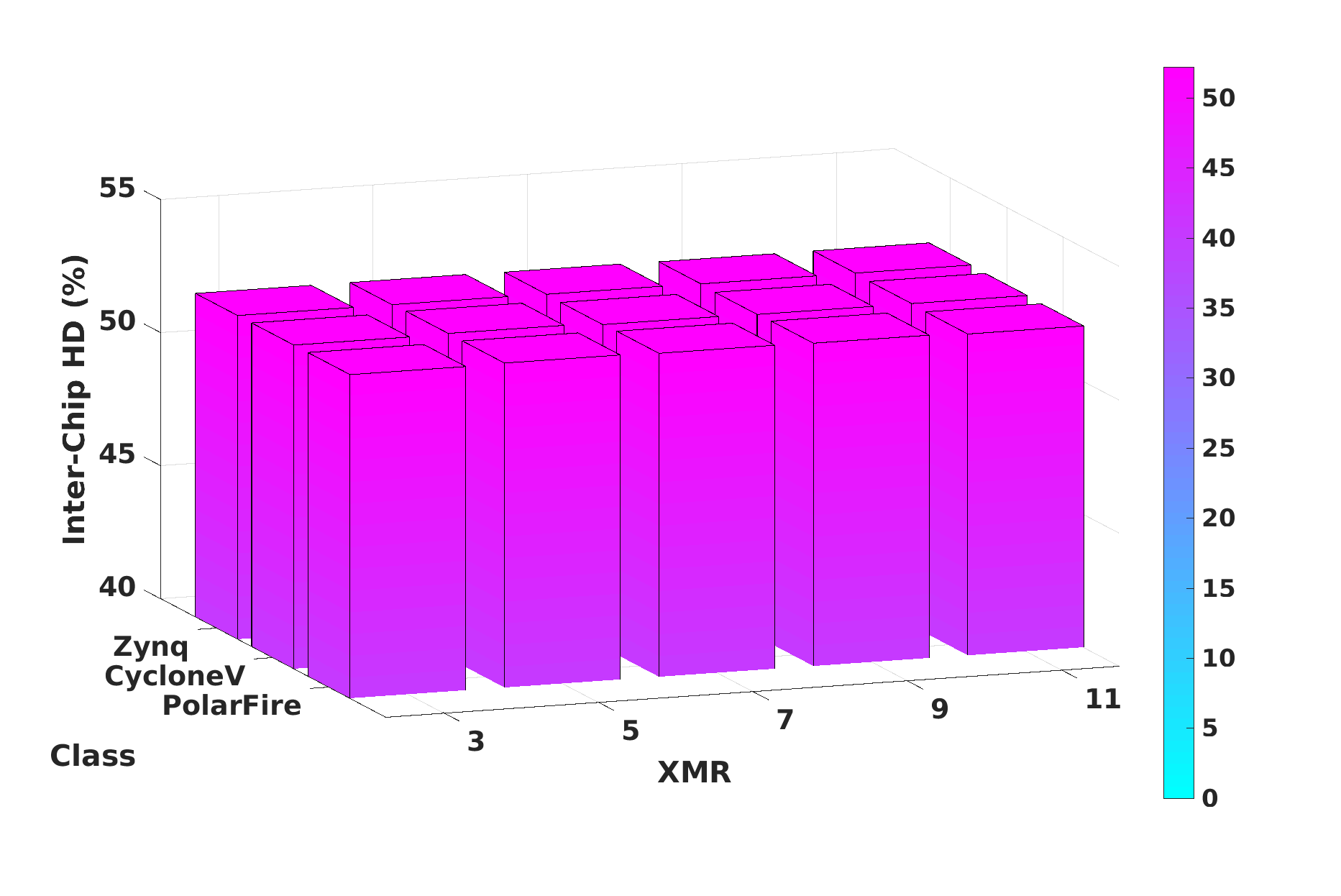}%
      \label{Fig:AlignedInterChipHD}}
   \caption{InterChip Hamming distance statistics for all device classes.}
   \label{Fig:InterchipHD_RawAligned}
   \vspace{-15pt}
\end{figure*}

In this section, we evaluate the super-strong bitstrings (\textbf{SBS}) generated by the 25 devices from each device class against statistical quality metrics including randomness, uniqueness and reliability. As indicated earlier, we regenerate the SBS using the same challenges across a set of 5 temperature conditions, and use the regenerated SBS to evaluate reliability. Randomness and uniqueness are evaluated using the enrollment-generated bitstrings only, which is possible when reliability statistics meet cryptographic standards of less than one bit flip in a million. To obtain statistically significant results, multiple challenges are used to generate bitstrings of size 5800 to more than 1.4 million bits per device depending on the requirements of the test. 

The bar graphs in Fig. \ref{Fig:EntropyMinEntropy} show the entropy and min-entropy statistical results for XMR values of 3 through 11 along the x-axis and for the three device classes along the y-axis. Entropy and min-entropy are computed using Eqs. \ref{Eq:Entropy} and \ref{Eq:MinEntropy}, respectively, where ${p_0}$ represents the fraction of bits that are '0', ${p_1}$ represents the fraction that are '1', and $p_{max}$ is the larger of ${p_0}$ and ${p_1}$. The best possible value of entropy and min-entropy is 1.0, which occurs when both fractions are 0.5.

The level of entropy across the device classes is nearly identical, where a slight decreasing trend is observable, from approximately 0.999 to 0.987, as XMR is increased from 3 to 11. Overall, the entropy results indicate very high levels exist across all three device classes, and the level is insensitive to the XMR level. The levels of min-entropy are again similar across the device classes but the sensitivity to XMR level is more noticeable, decreasing from approximately 0.93 at XMR 3 to 0.86 at XMR 11. However, despite the reduced levels, these results are similar to min-entropy levels published for other PUF architectures.

\begin{equation}
    H(x) = \sum\limits_{i=0}^{1}-(p_i \times log_2(p_i))
    \label{Eq:Entropy}
\end{equation}
\begin{equation}
    H_{\infty}(x) = -log_2(p_{max})
    \label{Eq:MinEntropy}
\end{equation}

\begin{figure*}[!t]
   \centering
   \subfloat[Probability of Failure]{\includegraphics[width=3.5in]{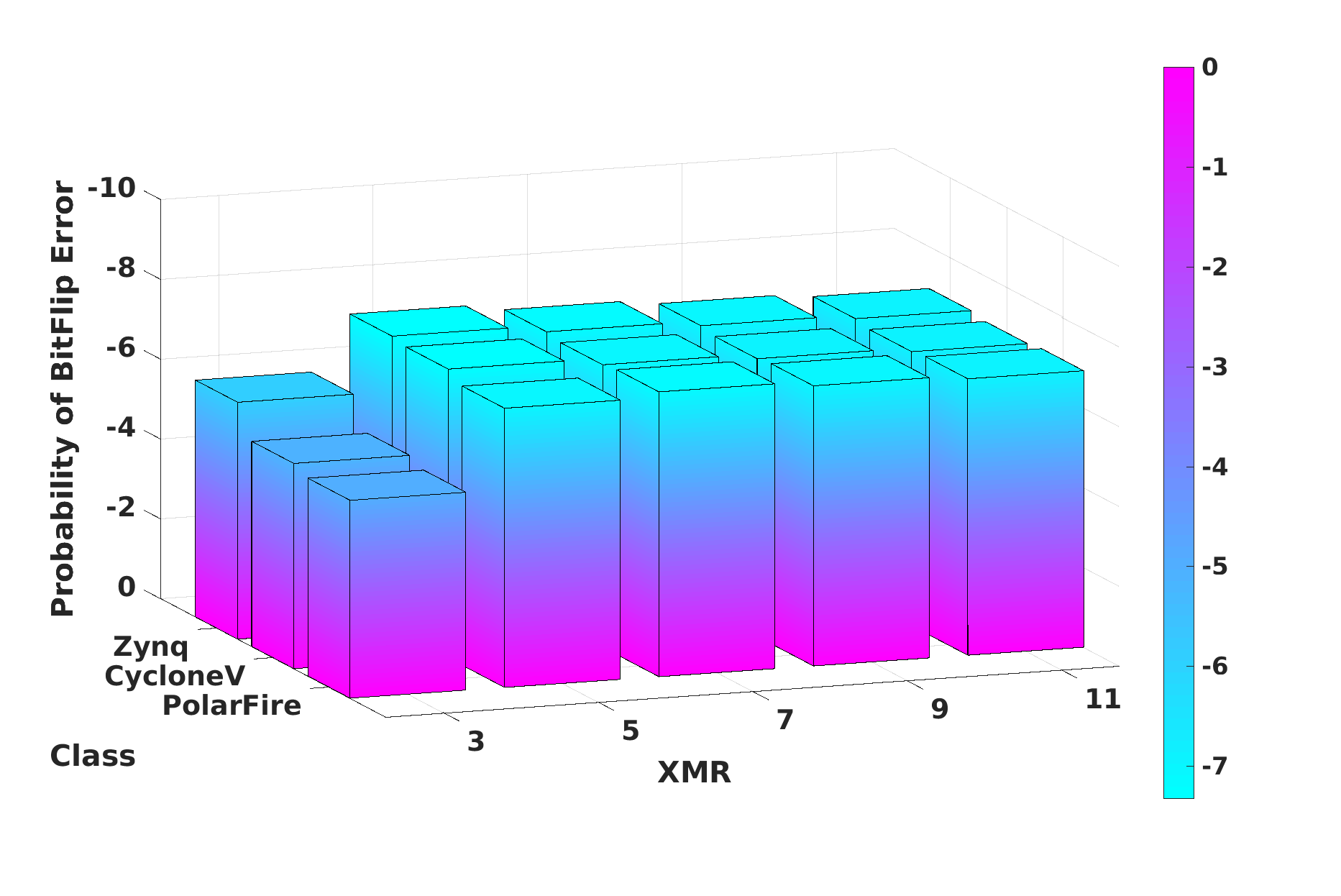}
      \label{Fig:POF}}
      \hfil
   \subfloat[Smallest Bitstring Size]{\includegraphics[width=3.5in]{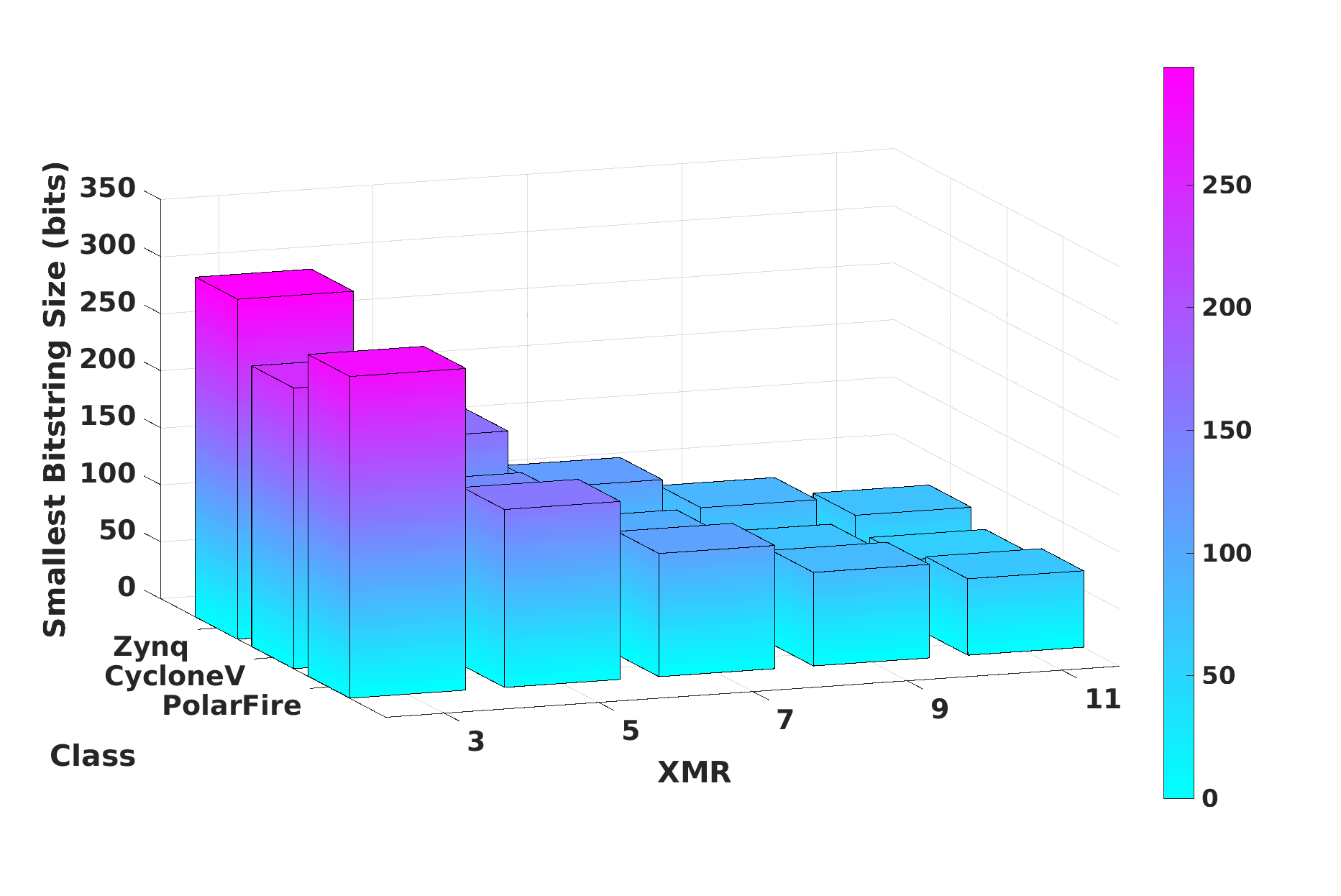}%
      \label{Fig:SmallestBitstringSize}}
   \caption{Probability of failure and smallest bitstring size statistics for all device classes.}
   \label{Fig:POF_and_SmallestBitstringSize}
   \vspace{-15pt}
\end{figure*}

The results of inter-chip Hamming distance (InterChip-HD) are shown in Fig. \ref{Fig:InterchipHD_RawAligned}, where we show the results using two different variants of the HD metric. Inter-HD measures the level of uniqueness across the bitstrings generated from the set of devices in each device class. Uniqueness is evaluated by pairing the enrollment bitstrings from each device class under all combinations (300 pairings with 25 devices) and then counting the number of bits that differ in each pairing. The best possible result occurs when the average number of bits that differ across all pairings is 50\%.

Both of the InterChip-HD and Aligned InterChip-HD metrics are computed using Eq. \ref{Eq:InterHD}. The difference is rooted in the selection of bit pairings that are used in the summation. For the traditional InterChip-HD results shown by the left bar graph, all bit pairings are used up to the length of the shorter bitstring. For the aligned analysis, a bit pairing is included only if the paths tested by the devices that generate the two bits in the pairing are the same. Although the InterChip-HD metric for the Aligned analysis more accurately reflects the uniqueness characteristics of the SiRF PUF, the super-strong bit selection processes associated with the Thresholding and XMR methods significantly reduce the number of bits that qualify. For example, only 45 bits on average are used per bitstring pairing for XMR 3, which decreases to only 3 bits on average for XMR 11. Therefore, the sample size for the Aligned analysis is much smaller. 

The bar graphs for both analyses show the average InterChip HD computed across all device bitstring pairings, where nearly ideal results of 50\% are achieved under the traditional analysis, and slightly larger values (approximately 52\%) are achieved under the Aligned analysis. There exists little or no distinction in the results for each of the device classes.

\begin{equation}
    \text{InterChip-HD}_{i,j} = \frac{\sum\limits_{k=1}^{\min(|bs_i|,|bs_j|)}{bs_{i,k} \oplus bs_{j,k}}}{\min(|bs_i|,|bs_j|)}
    \label{Eq:InterHD}
\end{equation}

The Probability of Failure (POF) results are shown in the left bar graph of Fig. \ref{Fig:POF_and_SmallestBitstringSize}, where failure refers to the occurrence of a bit-flip error(s). The reliability of the SiRF PUF in reproducing bitstrings without errors is measured in our experiments using data collected under 5 different temperatures, given by \{$-40~^\circ\text{C},  0~^\circ\text{C}, 25~^\circ\text{C}, 50~^\circ\text{C}, 85~^\circ\text{C}$\}. 

The POF results are derived from the intra-chip Hamming distance (IntraChip HD) metric given by Eq. \ref{Eq:IntraHD}, which counts the number of differences between a bitstring generated under nominal conditions and each of the bitstrings generated by the same device using the same challenges under different temperatures. The tuple ($i, n, j$) designates a bitstring pairing using the nominal bitstring $n$ and a bitstring generated under TV corner $j$ for device $i$. The total number of bit flip errors counted is converted to a POF by dividing the total number of bit flip errors by the total number of bits considered in the analysis. If no bit flip errors are detected in any device at any TV corner, we use the ratio of 1 over the number of bits evaluated as an upper bound approximation of reliability, which assumes one bit-flip occurred.

\begin{equation}
    \text{Intra-HD}_{i,n,j} = \sum\limits_{k=1}^{|bs_i|)}{bs_{i,n,k} \oplus bs_{i,j,k}}
    \label{Eq:IntraHD}
\end{equation}

The negative integer values shown along the z-axis of the bar graph in Fig. \ref{Fig:POF_and_SmallestBitstringSize} are the exponents of a value with base 10, so -6 corresponds to $10^{-6}$ or 1-in-a-million as the probability of failure. Bit-flip errors are counted separately for each of the 25 devices and then an overall metric is computed by taking the sum of bit-flip errors across all devices and dividing by the total number of bits inspected across all devices. In order to increase the significance of the results, a large set of challenges are applied to the devices. For example, the XMR 3 analysis inspected more than 37 million bits across all 25 devices in each device class, so the smallest value of any exponent is -7.58. 

Bit-flip errors occur in all device classes at XMR 3, where we see the reliability of the Zynq device class just meets the industry standard of $10^{-6}$, while for the Cyclone and PolarFire device classes, reliability is worse, and in the range of $10^{-5}$. However, for XMR 5, only 1 bit-flip is present in the Zynq and Cyclone analyses, and 2 in the PolarFire analysis with more than 17 million bits inspected. Although the reliability appears to degrade for XMRs 7, 9 and 11, it is due to the smaller numbers of bits inspected and is not due to bit-flip errors, in fact, none were observed at any of these XMR levels. These results indicate very high levels of reliability can be achieved for XMR values of 5 or above.

The right bar graph in Fig. \ref{Fig:POF_and_SmallestBitstringSize} shows the minimum number of bits generated using one iteration of the SiRF PUF algorithm, averaged across all devices in the class. The size of the SBS bitstring decreases as XMR is increased, as expected, because the number of bits used in the XMR redundancy scheme increases for larger XMR values. From the analysis presented in Section \ref{Subsection:EntropyAndTVNoise}, which shows higher levels of UC-TVN exist in the Cyclone device class, the primary penalty is shown here where the number of usable bits is smaller at each XMR level when compared with the Zynq and PolarFire device classes. Assuming XMR 5 is used due of reliability constraints, the average minimum number of bits for Zynq, Cyclone and PolarFire are 168, 137 and 158, respectively. Therefore, in all cases, two iterations of the SiRF PUF algorithm are needed to generate a 256-bit AES key at a XMR level of 5.

\begin{figure}[ht]
    \centering
    \includegraphics[width=3.4in,keepaspectratio=true]{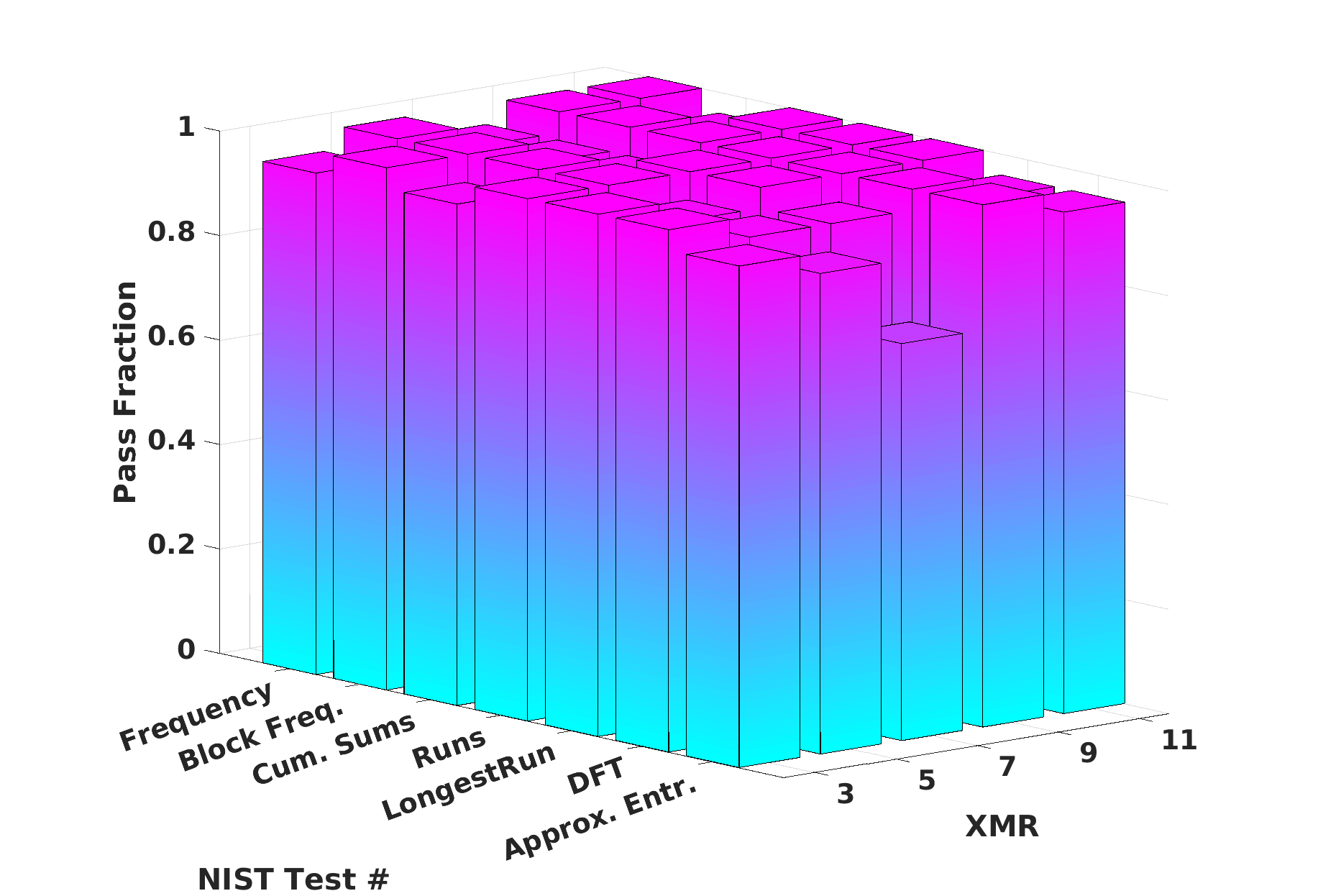}
    \caption{Zynq 7010 NIST statistcal results.}
    \label{fig:Zynq_NIST}
    \vspace{-10pt}
\end{figure}

\begin{figure}[ht]
    \centering
    \includegraphics[width=3.4in,keepaspectratio=true]{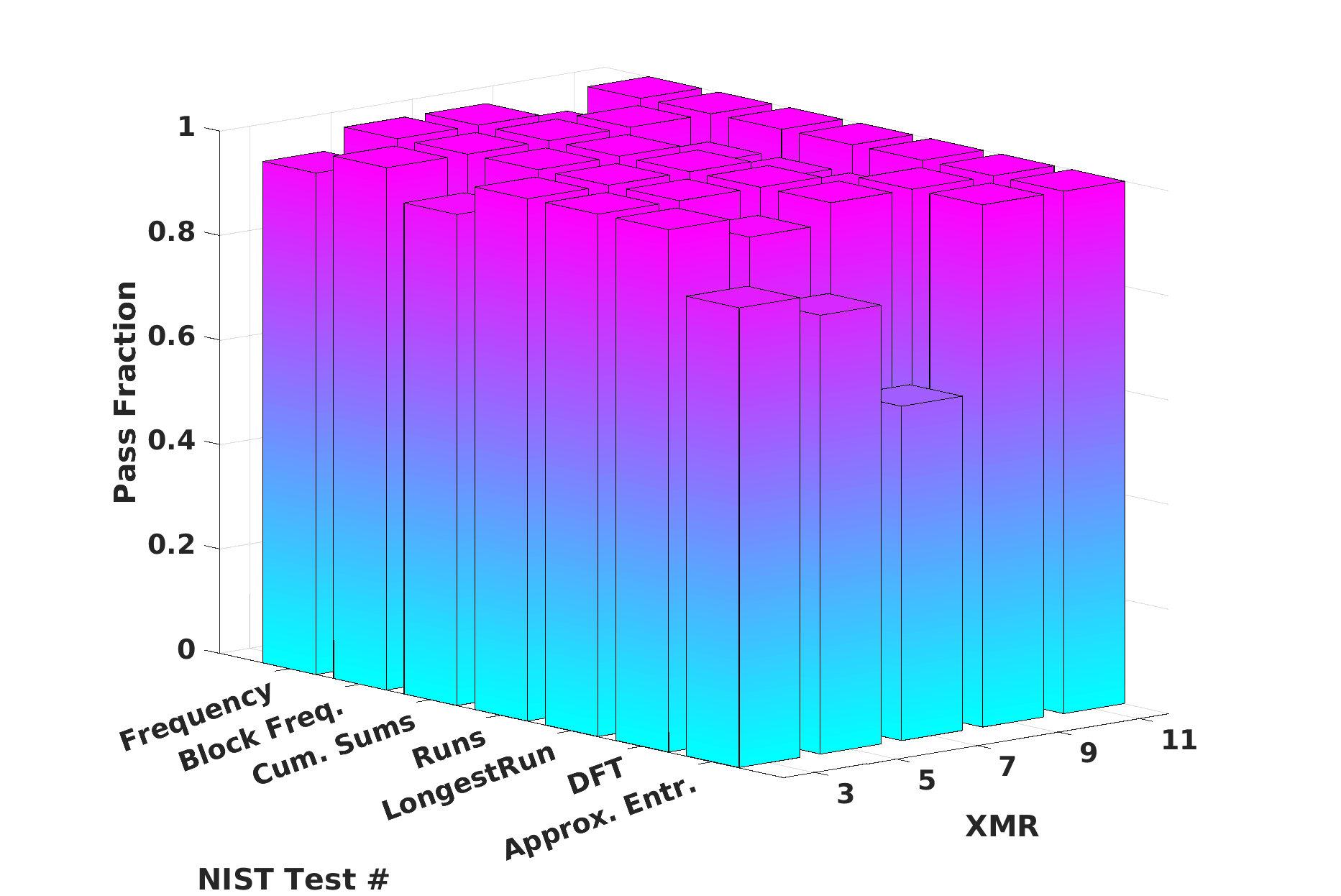}
    \caption{CycloneV NIST statistcal results.}
    \label{fig:CycloneV_NIST}
    \vspace{-10pt}
\end{figure}

\begin{figure}[ht]
    \centering
    \includegraphics[width=3.4in,keepaspectratio=true]{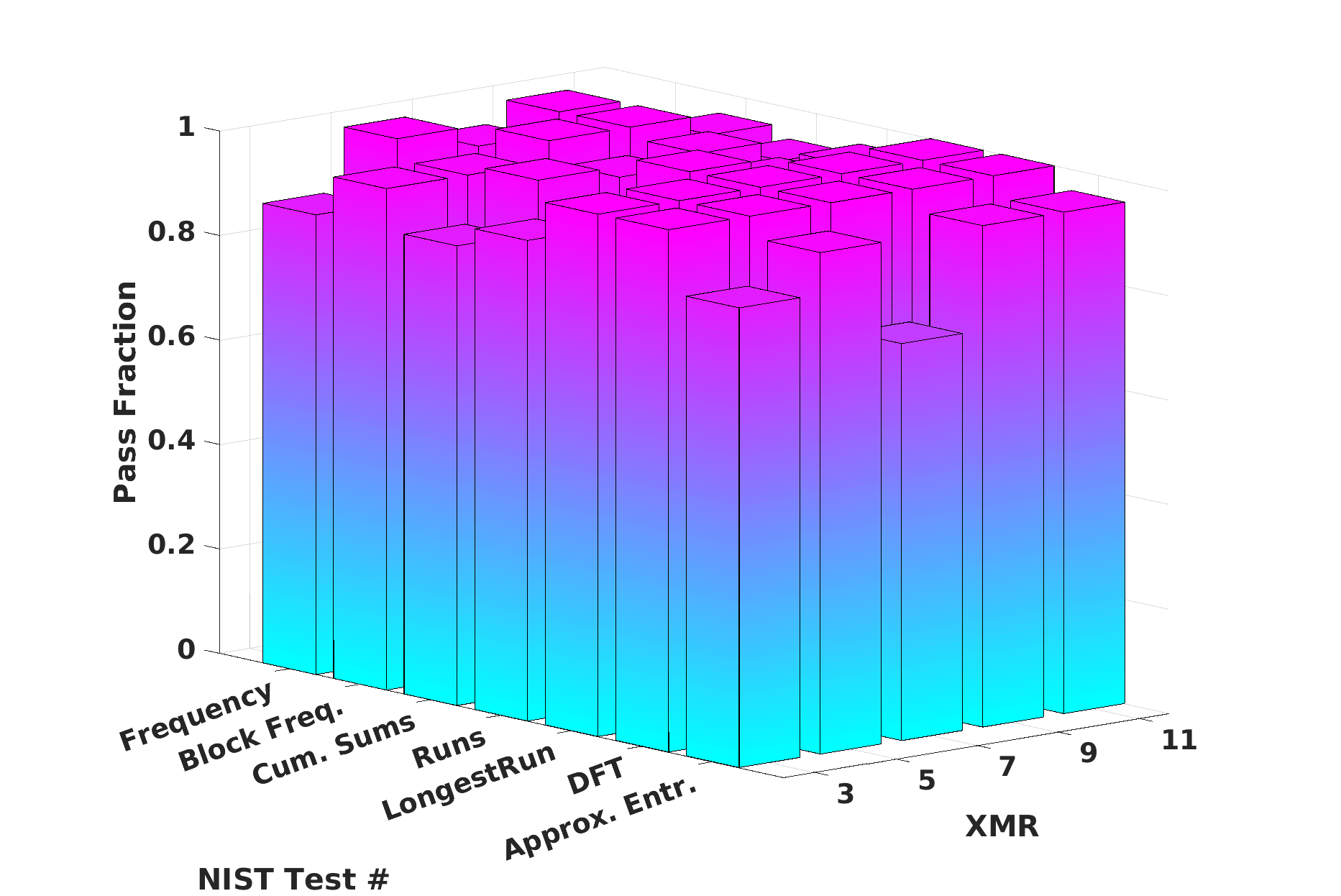}
    \caption{PolarFire NIST statistcal results.}
    \label{fig:PolarFire_NIST}
    \vspace{-10pt}
\end{figure}

The NIST results for each of the device classes are shown in Fig. \ref{fig:Zynq_NIST}, \ref{fig:CycloneV_NIST} and \ref{fig:PolarFire_NIST} \cite{NIST2010}. The size of the SBS subjected to NIST testing varies from 5800 for XMR 11 to nearly 30,000 for XMR 3, which enabled seven of the NIST statistical tests to be run. With a population of 25 devices, NIST requires that at least 23 of the devices pass each of the tests in order for the test to be considered passed overall. Therefore, bar heights below 0.92 indicate that 22 or fewer devices passed the test. The bar graphs indicate nearly all of the tests are passed for Zynq, except for one failure at XMR 7 for Approx. Entropy, where only 19 devices passed. For Cyclone, two additional fail cases are observed for the Approx. Entropy test, at XMR 3 and 5 with 22 and 21 devices passing, respectively. The worst case is again for XMR 7 with only 16 of the devices passing. PolarFire's results show three additional fail-by-1 cases for XMR 3 (Frequency, Cum. Sums and Approx. Entr.), but are otherwise similar to Zynq's results. Overall, despite the fail cases, the NIST results are generally very good, showing all three device classes are able to produce high quality bitstrings.

\section{Conclusions} \label{Section:Conclusions}

In this paper, we implement, test and compare the SiRF PUF architecture and the quality of the generated bitstrings on 25 copies of devices from Xilinx, Altera and Microsemi. The SiRF algorithm is run at room temperature to generate enrollment bitstrings, and then again with the devices placed in a temperature chamber and subjected to temperatures over the range from $-40~^\circ\text{C} to 85~^\circ\text{C}$, to regenerate the bitstrings. Statistical tests including entropy, min-entropy, inter-chip Hamming distance (HD), intra-chip HD, and tests from the NIST statistical test suite are used to evaluate the randomness, reliability and uniqueness characteristics of the bitstrings.

The results of our analysis show that all three devices produce high quality bitstrings suitable for cryptographic applications.
Overall, the SiRF PUF implemented on the Zynq platform performs slightly better than the Cyclone and PolarFire implementations, when assessed from a Entropy(signal)-to-(TV)noise perspective. Moreover, devices from the Cyclone class possess the highest level of TV-noise. However, the Zynq implementation is also the most mature and improvements are likely possible for the newer Cyclone and PolarFire implementations, which will be investigated in future work. High levels of statistical quality are reported for the bitstrings from all device classes, again, with Zynq performing slightly better. 
Another interesting artifact of the analysis, and a topic for future work, is the presence of distinguishing features in the delay distributions of devices fabricated in TSMC and UMC foundries.

\vspace{-20pt}
\section*{Acknowledgment}

We thank Raytheon for providing support and encouragement for the research results presented in this paper.

\vspace{-10pt}

\ifCLASSOPTIONcaptionsoff
  \newpage
\fi

\bibliographystyle{IEEEtran}
\bibliography{IEEEabrv,references}

\vspace{-40pt}

\begin{IEEEbiography}[{\includegraphics[width=0.63in,height=1.0in,clip,keepaspectratio]{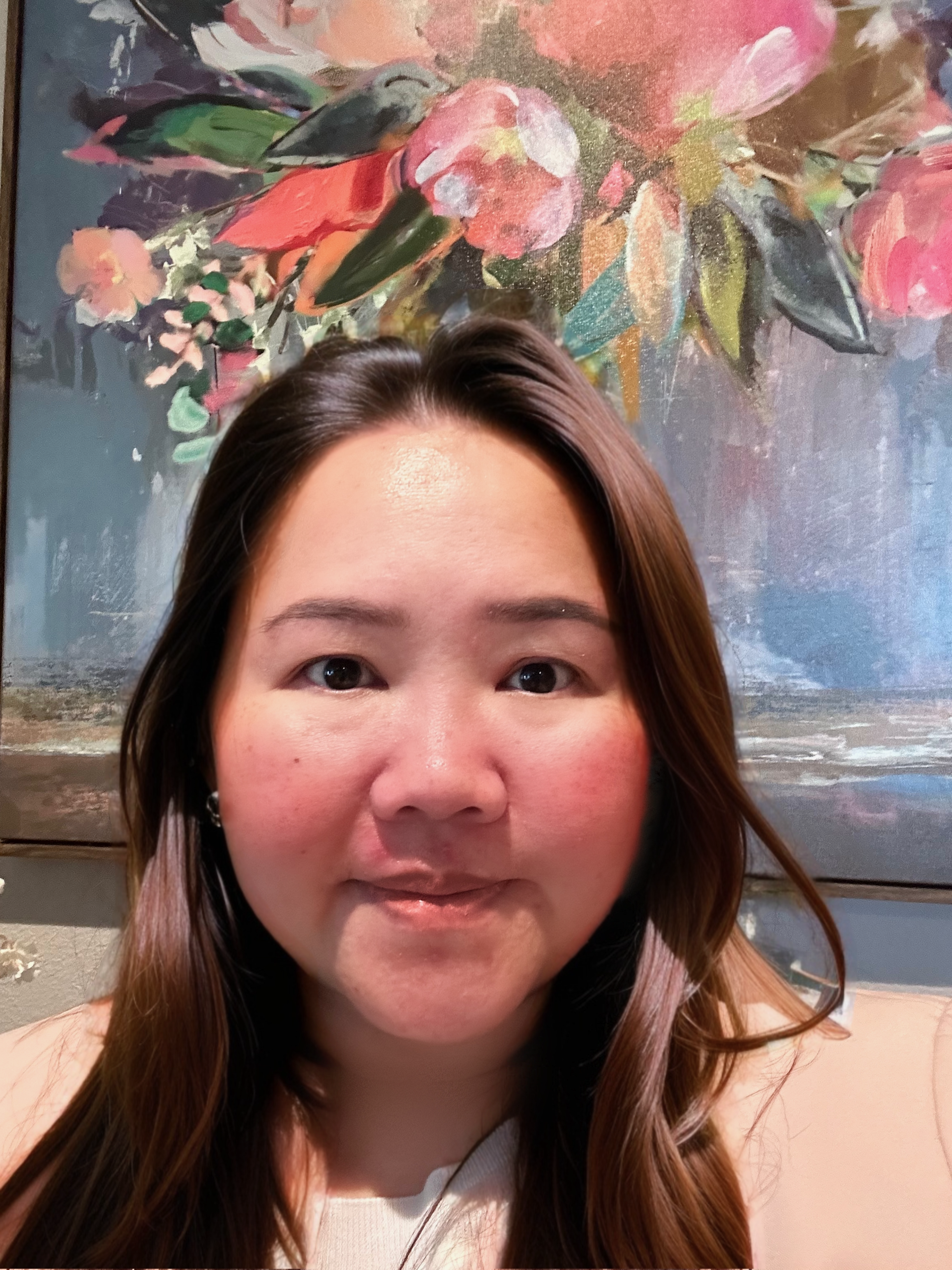}}]{Jenilee Jao}
is pursuing a Ph.D. in Computer Engineering with an emphasis on Hardware Security in the Electrical and Computer Engineering department at the University of New Mexico, where she also earned her B.S. and M.S. degrees.
\end{IEEEbiography}

\vspace{-40pt}

\begin{IEEEbiography}[{\includegraphics[width=0.63in,height=1.0in,clip,keepaspectratio]{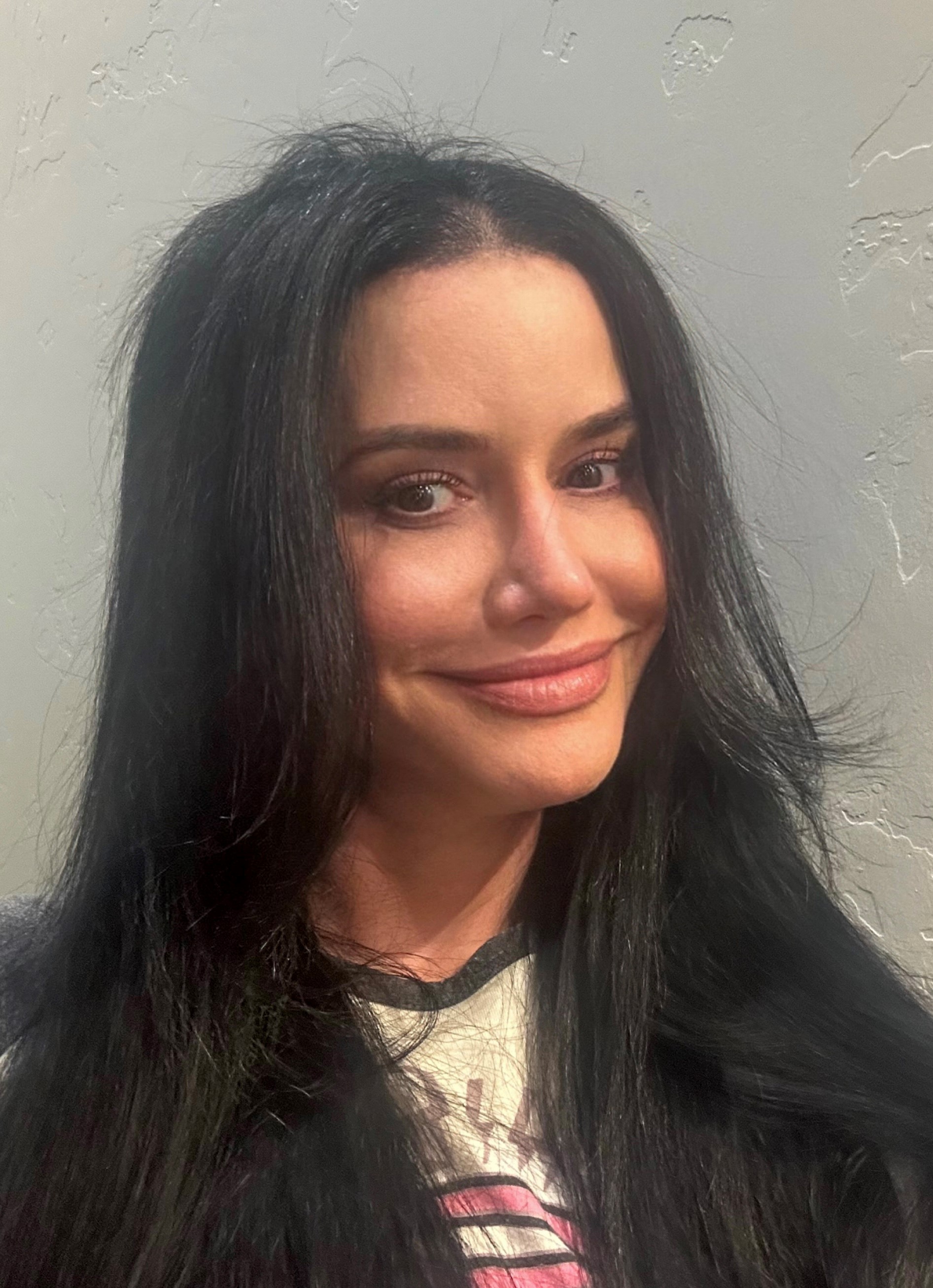}}]{Kristi Hoffman} is a Principal Tech Fellow at Collins Aerospace, Adjunct Professor at the University of New Mexico and a Certified Raytheon and Open Group Architect with over twenty years of experience.  Within not only the defense industry, but the electronics industry, she is recognized as an industry expert in, reverse engineering, anti-tamper and cyber. Kristi holds a Ph.D. from University of New Mexico.
\end{IEEEbiography}

\vspace{-40pt}

\begin{IEEEbiography}[{\includegraphics[width=0.63in,height=1.0in,clip,keepaspectratio]{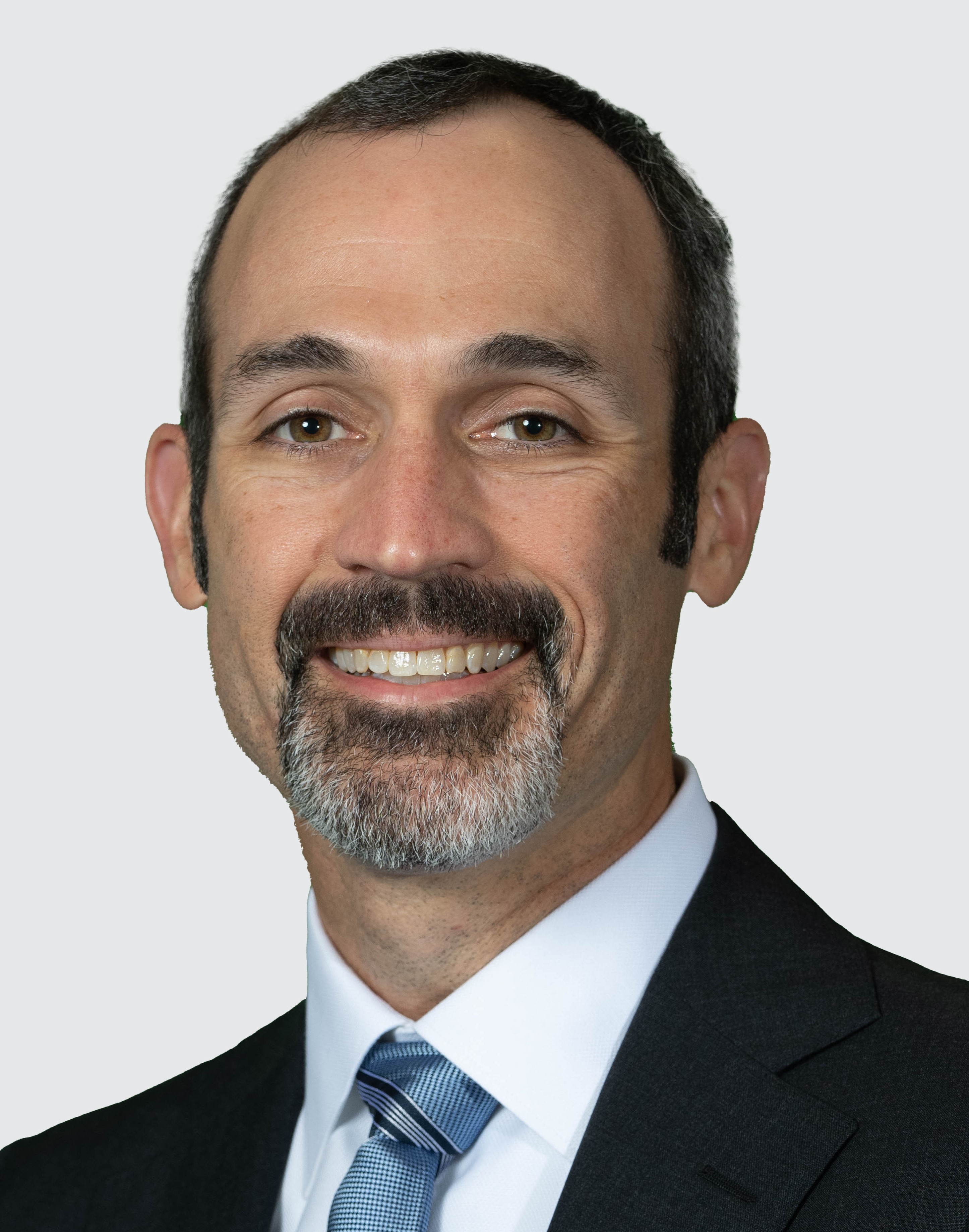}}]{Brent Emery}
is a Senior Principal Electrical Engineer at Collins Aerospace, an RTX company. He received a B.S in Electrical Engineering from the United States Military Academy at West Point and a M.S. in Electrical Engineering from the University of Texas at Dallas.
\end{IEEEbiography}

\vspace{-40pt}

\begin{IEEEbiography}[{\includegraphics[width=0.63in,height=1.0in,clip,keepaspectratio]{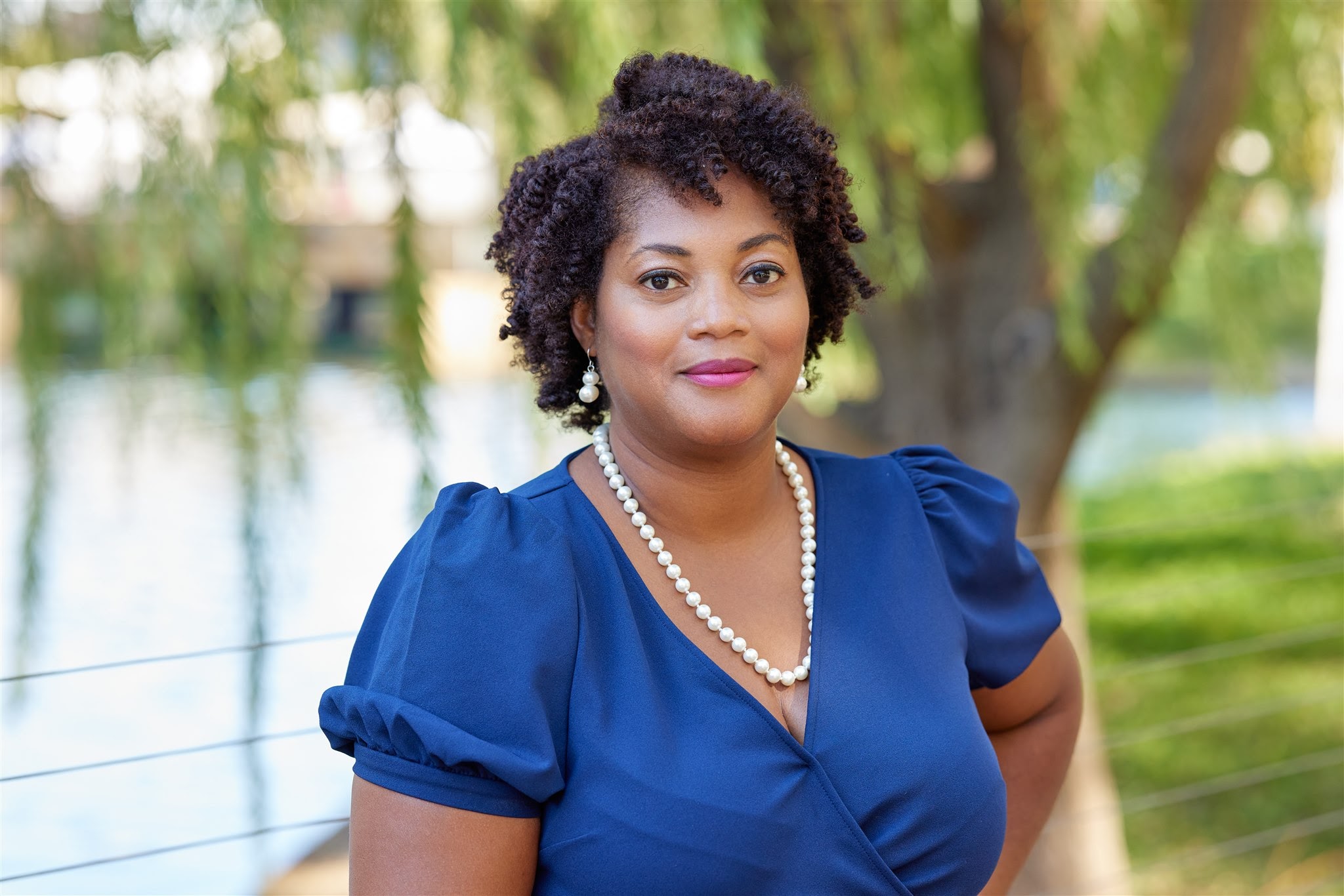}}]{Cheryl Reid}
is a Senior Principal Systems Engineer at Collins Aerospace, an RTX company. She received a B.S in Electrical Engineering New Jersey Institute of Technology and a M.S. in Electrical Engineering from Georgia Institute of Technology.
\end{IEEEbiography}

\vspace{-40pt}

\begin{IEEEbiography}[{\includegraphics[width=0.63in,height=1.0in,clip,keepaspectratio]{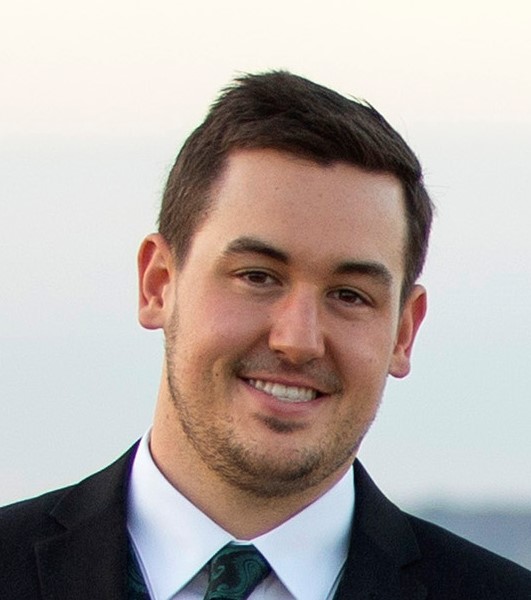}}]{Ryan Thomson}
is a Senior Software Engineer at Collins Aerospace, an RTX company. He received a B.S. in Computer Science from Stephen F. Austin State University.
\end{IEEEbiography}

\vspace{-30pt}

\begin{IEEEbiography}[{\includegraphics[width=0.63in,height=1.0in,clip,keepaspectratio]{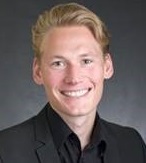}}]{Michael Thompson}
is a Principal System Security Engineer at Raytheon, an RTX company. He received a B.S. and M.S. in Electrical Engineering from Florida International University in Miami.
\end{IEEEbiography}

\vspace{-40pt}

\begin{IEEEbiography}[{\includegraphics[width=0.63in,height=1.0in,clip,keepaspectratio]{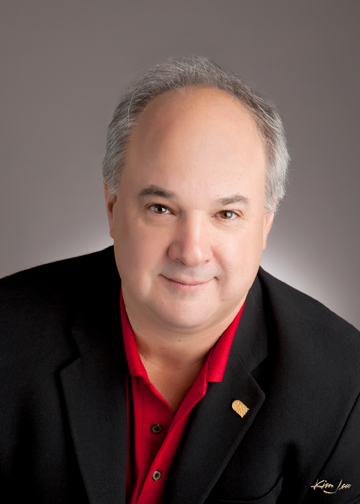}}]{Jim Plusquellic}
is a Professor in Electrical and Computer Engineering at the University of New Mexico. He received both his M.S. and Ph.D. degrees in Computer Science from the University of Pittsburgh. Professor Plusquellic received an "Outstanding Contribution Award" from IEEE Computer Society in 2012 and 2017 for co-founding and for his contributions to the Symposium on Hardware-Oriented Security and Trust (HOST).
\end{IEEEbiography}

\end{document}